\documentclass[10pt,twocolumn,twocolappendix]{aastex63}

\usepackage{chngcntr}

\usepackage{amsmath}
\usepackage{soul}
\newcommand{\ms}{$\rm m\,s^{-1}$}
\newcommand{\kms}{$\rm km\,s^{-1}$}

\newcommand{\vsini}{\textit{v}\,\textnormal{sin}\,\textit{i}}
\newcommand{\logg}{log\,$g$}
\newcommand{\htwoo}{$\rm H_{2}O$}
\newcommand{\teff}{$T_{\rm eff}$}

\received{February 10, 2023}
\revised{March 17, 2023}
\accepted{March 17, 2023}
\submitjournal{ApJ}

\graphicspath{{./}{figures/}}
\shorttitle{DI\,Tau\,AB}
\shortauthors{Tang et al.}

\begin{document}

\title{
    Star-Crossed Lovers DI\,Tau\,A and B: Orbit Characterization and Physical Properties Determination
    }


\author[0000-0003-4247-1401]{Shih-Yun Tang}
    \affiliation{Lowell Observatory, 1400 West Mars Hill Road, Flagstaff, AZ 86001, USA}
    \affiliation{Department of Physics and Astronomy, Rice University, 6100 Main Street, Houston, TX 77005, USA}
    \affiliation{Department of Astronomy and Planetary Science, Northern Arizona University, Flagstaff, AZ 86011, USA}
    
\author[0000-0002-0848-6960]{Asa G. Stahl}
    \affiliation{Department of Physics and Astronomy, Rice University, 6100 Main Street, Houston, TX 77005, USA}

\author[0000-0001-7998-226X]{L. Prato}
    \affiliation{Lowell Observatory, 1400 West Mars Hill Road, Flagstaff, AZ 86001, USA}
    \affiliation{Department of Astronomy and Planetary Science, Northern Arizona University, Flagstaff, AZ 86011, USA}

\author[0000-0001-5415-9189]{G. H. Schaefer}
    \affiliation{The CHARA Array of Georgia State University, Mount Wilson Observatory, Mount Wilson, CA 91023, USA}
    
\author[0000-0002-8828-6386]{Christopher M. Johns-Krull}
    \affiliation{Department of Physics and Astronomy, Rice University, 6100 Main Street, Houston, TX 77005, USA}

\author[0000-0001-5306-6220]{Brian A. Skiff}
    \affiliation{Lowell Observatory, 1400 West Mars Hill Road, Flagstaff, AZ 86001, USA}
    
\author[0000-0002-5627-5471]{Charles A. Beichman}
    \affiliation{Infrared Processing and Analysis Center, California Institute of Technology, 1200 E. California Boulevard, Pasadena, CA 91125, USA}
    \affiliation{NASA Exoplanet Science Institute, Pasadena, CA 91125, USA}
    
\author[0000-0002-6879-3030]{Taichi Uyama}
    \affiliation{Infrared Processing and Analysis Center, California Institute of Technology, 1200 E. California Boulevard, Pasadena, CA 91125, USA}
    \affiliation{NASA Exoplanet Science Institute, Pasadena, CA 91125, USA}
    \affiliation{National Astronomical Observatory of Japan, 2-21-1 Osawa, Mitaka, Tokyo 181-8588, Japan}

\email{sytang@lowell.edu, Asa.Stahl@rice.edu}

\begin{abstract}

The stellar companion to the weak-line T Tauri star DI\,Tau\,A was first discovered by the lunar occultation technique in 1989 and was subsequently confirmed by a speckle imaging observation in 1991. It has not been detected since, despite being targeted by five different studies that used a variety of methods and spanned more than 20 years. Here, we report the serendipitous rediscovery of DI\,Tau\,B during our Young Exoplanets Spectroscopic Survey (YESS). Using radial velocity data from YESS spanning 17 years, new adaptive optics observations from Keck\,II, and a variety of other data from the literature, we derive a preliminary orbital solution for the system that effectively explains the detection and (almost all of the) non-detection history of DI\,Tau\,B. We estimate the dynamical masses of both components, finding that the large mass difference (q $\sim$0.17) and long orbital period ($\gtrsim$35 years) make DI\,Tau system a noteworthy and valuable addition to studies of stellar evolution and pre-main-sequence models. With a long orbital period and a small flux ratio (f2/f1) between DI\,Tau\,A and B, additional measurements are needed for a better comparison between these observational results and pre-main-sequence models. Finally, we report an average surface magnetic field strength ($\bar B$) for DI\,Tau\,A, of $\sim$0.55 kG, which is unusually low in the context of young active stars. 
\end{abstract}



\section{Introduction}\label{sec:intro}

A star's mass, chemical composition, and magnetic field strength determine how its physical properties evolve over time. At pre-main-sequence (PMS) stages, there are significant differences in the evolutionary tracks predicted by different sets of models \citep[e.g.,][]{hillenbrand2004,simon2019}. Mapping a binary star system's astrometric orbit combined with radial velocity (RV) data provides a way to measure the dynamical masses of the stellar components and test the predictions of evolutionary models \citep[e.g.,][]{rizzuto2020a,biller2022}. At the distance of the Taurus star-forming region \citep[$\sim$140pc,][]{torres2009a,roccatagliata2020}, a face-on binary in a circular orbit with a total mass of 1 $M_\odot$ and a period of 10 years would have a maximum projected separation of $\sim$33 milliarcseconds (mas) on the plane of the sky. Binaries in nearby star-forming regions that can be resolved through high-resolution imaging techniques, such as speckle interferometry or the use of adaptive optics (AO), have periods typically longer than a decade \citep[e.g.,][]{schaefer2006}. Taken together, these conditions make mapping out the orbits of young binaries difficult. Longer-period systems in particular are more easily overlooked by RV surveys, as their RV variations can be small. Young visual binaries tend to have orbital eccentricities larger than 0.2 \citep[e.g.,][]{schaefer2012,schaefer2014,schaefer2020a,allen2017,rizzuto2016,rizzuto2020a,czekala2021,zuniga-fernandez2021}. Given that higher eccentricity systems spend most of their time far from periastron, they are even more likely to be missed by RV surveys and, thus, potentially leave a hole in the binary demographic.

DI\,Tau (HBC~39) is a PMS star with an age of about 1.6 Myr \citep{herczeg2014} and a spectral type of M0 \citep{nguyen2012} located at a distance of $\sim$137 pc \citep{katz2022,bailer-jones2021} in the Taurus-Auriga star-forming region (Table~\ref{tab:basic}). DI\,Tau\,B was first discovered by \citet{chen1990} using the lunar occultation technique at the Wyoming Infrared Observatory on 1989 August 24. This discovery was later confirmed by \citet{ghez1993a} with speckle imaging observed on 1991 October 18 at the Hale 5 m Telescope of Palomar Observatory. The results from these two studies are included in Table~\ref{tab:relA}. 

Follow-up observations failed to recover DI\,Tau\,B. \citet{simon1996} observed DI\,Tau twice (on 1993 September 26 and 1993 October 25) with the Hubble Space Telescope's (HST) Fine Guidance Sensors (FGS) and reported no detection. Using the near-infrared (NIR) Coronagraph Imager with Adaptive Optics (CIAO) at the Subaru telescope on 2004 January 11, \citet{itoh2008} described a candidate companion to DI\,Tau\,A with a separation of 5\farcs18 and $\Delta K$ of $\sim$10 mag, later shown to be an unrelated background object based on statistical studies of proper motion \citep{daemgen2015}. \citet{kraus2011} observed DI\,Tau on 2008 December 23 using Keck\,II NIRC2 with a non-redundant mask and also reported a non-detection on DI\,Tau\,B.
The last imaging that we know of for DI\,Tau prior to the observations described here was published in \citet{schaefer2014} using the Keck\,II NIRC2 with AO on 2013 January 27 and again resulted in a non-detection.

Beyond the detections and non-detections of DI\,Tau\,B, this system poses another uncertainty\,---\,is DI\,Tau a Weak-line T\,Tauri Star (WTTS) or a Classical T\,Tauri Star (CTTS)? Some studies have classified DI\,Tau as a CTTS (disk-bearing PMS) because it shows mid-IR excess \citep[e.g.,][]{kenyon1995,sullivan2022}; however, others classify DI\,Tau as a WTTS because it exhibits little to no veiling in optical spectra \citep[e.g.,][]{nguyen2012,herczeg2014}.

In this study, we identify the elusive companion of DI\,Tau\,A with RV data from our Young Exoplanets Spectroscopic Survey (YESS) \citep{prato2008,crockett2012,johns-krull2016}. The YESS program was initiated in 2004 to search for stellar and substellar companions to young active stars, mostly T Tauri stars (TTSs), using RVs from optical and NIR wavelength regions. Here, we use optical and NIR RVs data from YESS spanning 17 years alongside new Keck\,II NIRC2 AO images and other literature data to determine an orbital solution and fundamental stellar parameters for the DI\,Tau system. We also show that DI\,Tau is likely a WTTS. In what follows, Section~\ref{sec:data} describes the observation and data reduction of the RVs, AO imaging, and photometry used in this study. In Section \ref{sec:orbit_fit}, we describe a Keplerian orbital fit to the combined RV, imaging, and lunar occultation data. The measurement of DI\,Tau\,A's mean surface magnetic field strength is presented in Section~\ref{sec:B}. In Section~\ref{sec:photo}, we evaluate DI\,Tau's $V$ band photometric variability over a decade-long timescale and describe a two-component SED fit to the resolved and unresolved photometry. Finally, in Section~\ref{sec:discussion}, we discuss the non-detections of DI\,Tau\,B in the literature, evaluate the lack of evidence for a circumstellar disk around  DI\,Tau\,A, and compare our dynamical mass estimates with those predicted by theoretical evolutionary models. A brief summary of this study is given in Section~\ref{sec:summary}.

\begin{deluxetable}{l C c}
\tablecaption{
    DI\,Tau Basic Properties\label{tab:basic}}
\tabletypesize{\scriptsize}
\tablehead{
    \colhead{\hspace{15pt}Parameter}\hspace{20pt}   &
    \colhead{\hspace{20pt}Value}\hspace{20pt}   &
    \colhead{\hspace{20pt}Source$^a$}\hspace{20pt}
     }
\startdata 
\multicolumn{3}{c}{Astrometry}  \\
\hline
R.A. (J2016.0)                              & 67.4270079            & Gaia EDR3 \\
Decl. (J2016.0)                             & 26.5468834            & Gaia EDR3 \\
$\mu_\alpha \cos{\delta}$ (mas yr$^{-1}$)   &  8.007  \pm 0.049     & Gaia EDR3 \\
$\mu_\delta$ (mas yr$^{-1}$)                &-21.771  \pm 0.037     & Gaia EDR3 \\
$\varpi$ (mas)                              &  7.2690 \pm 0.0467    & Gaia EDR3 \\
Distance (pc)                   & 137.35     \pm 0.86     & \citet{bailer-jones2021} \\
Mean RV (\kms)                              & 13.16	  \pm 3.56      & Gaia DR3 \\
\hline
\multicolumn{3}{c}{Photometry}\\
\hline
$G_{\rm BP}$ (mag)  & 13.086 \pm 0.0051   & Gaia DR3  \\
$G$ (mag)           & 11.959 \pm 0.0030   & Gaia DR3  \\
$G_{\rm RP}$ (mag)  & 10.903 \pm 0.0048   & Gaia DR3  \\
$J$ (mag)           &  9.323 \pm 0.026    & 2MASS     \\
$H$ (mag)           &  8.599 \pm 0.024    & 2MASS     \\
$Ks$ (mag)          &  8.391 \pm 0.020    & 2MASS     \\
$W1$ (mag)          &  8.267 \pm 0.023    & WISE      \\
$W2$ (mag)          &  8.239 \pm 0.018    & WISE      \\
$W3$ (mag)          &  8.040 \pm 0.024    & WISE      \\
IRAC 36 (mag)      &  8.296 \pm 0.058   & Spitzer     \\
IRAC 45 (mag)      &  8.410 \pm 0.053   & Spitzer     \\
IRAC 58 (mag)      &  8.135 \pm 0.051    & Spitzer    \\
IRAC 80 (mag)      &  8.069 \pm 0.051    & Spitzer    \\
\hline \\[-10pt]
\enddata
\tablecomments{$^a$ Gaia EDR3: \citet{lindegren2021a,riello2021}; Gaia DR3: \citet{katz2022}; 
    2MASS: \citet{skrutskie2006}; WISE: \citet{cutri2012}; Spitzer: \citet{evans2003}.
    }
\end{deluxetable}

\begin{deluxetable*}{c c LL L ll}
\tablecaption{DI\,Tau Relative Astrometry Properties\label{tab:relA}
             }
\tabletypesize{\scriptsize}
\tablehead{
    \colhead{Epoch (UT)}    & \colhead{Julian Year}        &
    \colhead{$\rho$}        & \colhead{P.A.}        &
    \colhead{Flux Ratio}    & \colhead{Filter}      & \colhead{Reference}  \\
%
	\colhead{(yyyy-mm-dd)}  & \colhead{}            & 
	\colhead{(mas)}         & \colhead{(deg)}       &
	\colhead{f2/f1}         & \colhead{}            & \colhead{}            \\
	\colhead{(1)} & \colhead{(2)} & \colhead{(3)} & 
	\colhead{(4)} & \colhead{(5)} & \colhead{(6)} & \colhead{(7)}
     }
\startdata 
\hline
\multicolumn{7}{c}{Image}   \\
\hline
1991-10-18 & 1991.7945  & 120 \pm 10        & 294 \pm 4         & 0.13 \pm 0.01     & $K$                 & \citet{ghez1993a}  \\
2022-02-13 & 2022.1184  & 75.84 \pm 2.49    & 292.31 \pm 1.88   & 0.203 \pm 0.022  & K$_{\rm cont}$    & This study        \\
2022-10-19 & 2022.7980  & 79.98 \pm 2.10    & 289.64 \pm 1.51   & 0.159 \pm 0.032  & J$_{\rm cont}$    & This study        \\
           &            &                   &                   & 0.118 \pm 0.015  & H$_{\rm cont}$    & This study        \\
           &            &                   &                   & 0.188 \pm 0.017  & K$_{\rm cont}$    & This study        \\
           &            &                   &                   & 0.244 \pm 0.021  & $L'$              & This study        \\
2022-12-29 & 2022.9918  & 80.78 \pm 2.65    & 290.03 \pm 1.89   & 0.145 \pm 0.010  & J$_{\rm cont}$    & This study        \\
           &            &                   &                   & 0.127 \pm 0.011  & H$_{\rm cont}$    & This study        \\
           &            &                   &                   & 0.183 \pm 0.011  & K$_{\rm cont}$    & This study        \\
           &            &                   &                   & 0.232 \pm 0.009  & $L'$              & This study        \\
\hline
\multicolumn{7}{c}{1-D Lunar Occultation$^{a}$}   \\
\hline
1989-08-24 & 1989.6441  & 72.1 \pm 0.7      & 257               & 0.13 \pm 0.02     & $K$                 & \citet{chen1990} \\
\hline \\[-9pt]
\enddata
\tablecomments{
    $^a$ for the 1-D lunar occultation observation, P.A. indicate is the direction of the lunar occultation, and $\rho$ is the projected separation along the P.A..
    }
\end{deluxetable*}


\section{Observations and Data Reduction}\label{sec:data}

Three different types of data are used in this study: spectroscopy, AO imaging, and photometry. In the following subsections, we introduce each by describing the observations first and then the reduction process.

\subsection{Spectroscopy and Radial Velocities}\label{sec:rv}
\subsubsection{Optical}\label{sec:optrv}

Optical spectra were obtained with the Robert G. Tull Coud\'e Spectrograph \citep{tull1995} mounted on the 2.7 m Harlan J. Smith telescope at the McDonald Observatory. A detailed description on the setup of these observations can be found in \citet{prato2008,mahmud2011,crockett2012}. Here, we only give a short summary and provide information related to our DI\,Tau observations. In total, we obtained 22 spectra of DI\,Tau between January 2005 and January 2022, spanning about 17 years (Table~\ref{tab:rv}). Typical exposure times were $\sim$2500\,s, but ranged from 1600\,s to 3600\,s depending on the sky conditions. Each observation used a 1\farcs2 slit width, achieving a resolving power R~$\equiv\lambda/\Delta \lambda \sim$60\,000 and covering 3986 to 9952{\r A}. Wavelength calibration relied on the Th-Ar lamp exposures taken before and after each DI\,Tau observation. 

We reduced the optical spectra with custom IDL code built on the procedures of \citet{valenti1994} and \citet{hinkle2000}. Spectra were bias subtracted, flat fielded, and corrected for scattered light. We then proceeded with optimal spectral extraction. The effect of the blaze on each order's extracted spectrum is removed in two steps: first, by dividing by the extracted spectrum of the flat lamp, and second, by dividing by a second-order polynomial fit to the intensity of each order. For wavelength calibration we used approximately 1800 Th-Ar lines observed with an internal comparison lamp exposures and fit with a two-dimensional polynomial function in pixel space. The dispersion solution was determined with a two-dimensional polynomial fit to the line locations in wavelength space.

For our RV analysis we used observations of six standard stars (HD\,4628, 107\,Psc, $\tau$\,Ceti, HD\,88371, HD\,80367, and HD\,65277) for absolute RV uncertainty calibration. This external precision was combined with the measured internal RV precision (i.e., the standard deviation of RVs obtained from different echelle orders) for each target star to obtain the total uncertainty in the absolute RV measurement, similar to the procedure described in \citet{stahl2022}. Overall, our absolute RVs are accurate to a median level of 0.531 \kms, adequate for characterizing long-period binary orbits.

\begin{deluxetable}{c C CC}
\tablecaption{DI\,Tau Radial Velocities\label{tab:rv}
             }
\tabletypesize{\scriptsize}
\tablehead{
    \colhead{\hspace{20pt}UT}\hspace{20pt}                  & 
    \colhead{\hspace{10pt}JD$-$2450000}\hspace{10pt}        &
    \colhead{\hspace{15pt}RV}\hspace{15pt}                  &
    \colhead{\hspace{15pt}$\sigma_{\rm RV}$}\hspace{15pt}   \\
	\colhead{(yyyy-mm-dd)}    & \colhead{(days)}          & 
	\multicolumn{2}{c}{(\kms)}                      \\
	 \cline{3-4}
	 \colhead{(1)} & \colhead{(2)} & \colhead{(3)} & 
	 \colhead{(4)}
     }
\startdata 
\multicolumn{4}{c}{McDonald Optical RV}\\
\hline
2005-01-04 & 3374.63557 & 15.309 & 0.526 \\
2005-11-23 & 3697.94388 & 15.335 & 0.480 \\
2012-01-04 & 5930.73613 & 14.213 & 0.502 \\
2012-01-07 & 5933.70076 & 14.528 & 0.503 \\
2012-11-24 & 6255.80512 & 13.887 & 0.506 \\
2012-11-25 & 6256.75581 & 13.841 & 0.503 \\
2012-11-26 & 6257.75658 & 13.608 & 0.510 \\
2013-11-10 & 6606.78132 & 14.101 & 0.495 \\
2015-08-30 & 7264.97482 & 13.219 & 0.567 \\
2015-08-31 & 7265.96691 & 13.357 & 0.510 \\
2015-09-01 & 7266.95489 & 13.659 & 0.525 \\
2015-09-02 & 7267.95900 & 13.683 & 0.564 \\
2021-11-12 & 9530.78801 & 16.298 & 0.473 \\
2021-11-13 & 9531.78707 & 16.606 & 0.474 \\
2021-11-14 & 9532.78109 & 16.309 & 0.490 \\
2021-11-15 & 9533.81662 & 16.289 & 0.462 \\
2021-11-16 & 9534.77699 & 16.244 & 0.472 \\
2021-11-17 & 9535.76688 & 16.327 & 0.471 \\
2021-12-30 & 9578.70943 & 16.432 & 0.454 \\
2021-12-31 & 9579.68523 & 16.307 & 0.453 \\
2022-01-03 & 9582.72080 & 16.340 & 0.454 \\
2022-01-04 & 9583.64689 & 16.414 & 0.453 \\
2022-01-05 & 9584.71378 & 16.339 & 0.471 \\
2022-11-13 & 9896.78786 & 16.538 & 0.482 \\
2022-11-14 & 9897.83955 & 16.545 & 0.568 \\
2022-11-18 & 9901.77441 & 16.890 & 0.538 \\
\hline
\multicolumn{4}{c}{IGRINS NIR RV}\\
\hline
2015-10-29 & 7324.80190 & 13.635 & 0.055 \\
2016-10-04 & 7665.97106 & 13.449 & 0.056 \\
2016-12-08 & 7730.89895 & 13.294 & 0.056 \\
2017-09-29 & 8025.91475 & 15.410 & 0.057 \\
2017-11-23 & 8080.84341 & 15.649 & 0.055 \\
\hline \\[-9pt]
\enddata
\tablecomments{
    This table is available in machine-readable form.
    }
\end{deluxetable} 

\subsubsection{Near-Infrared}\label{sec:nirrv}

NIR spectra were taken using the high resolution Immersion GRating INfrared Spectrometer \citep[IGRINS,][]{yuk2010,park2014,mace_2016_56434,levine2018a}. IGRINS is a cross-dispersed echelle spectrograph that simultaneously covers the full $H$ (1.49--1.80 \micron{} split into 25 orders) and $K$ bands (1.96--2.46 \micron{} split into 22 orders). With no moving parts, the spectrograph's fixed 0\farcs8 width slit delivers a R$\sim$45\,000. IGRINS has been deployed at the McDonald Observatory's 2.7 m Harlan J. Smith Telescope, the 4.3 m Lowell Discovery Telescope (LDT, formerly the Discovery Channel Telescope, DCT), and the Gemini South telescope. 

DI\,Tau was observed once with IGRINS at McDonald Observatory on 2015 October 10 and four more times at the LDT in 2016 and 2017 (Table~\ref{tab:rv}). Observations of DI\,Tau were taken with either an AB or ABBA nodding sequence. The AB nodding sequences used a longer exposure time of 300\,s, while the ABBA sequences used shorter exposures of 100\,s, except for the observation at McDonald, which had an exposure time of 400\,s with ABBA nodding.

IGRINS spectral reduction was done using the IGRINS pipeline package version 2.2.0 \citep[plp\,v2.2.0\footnote{\url{https://github.com/igrins/plp}};][]{jae_joon_lee_2017_845059}, and the associated NIR RVs were obtained using the python package \texttt{IGRINS RV}\footnote{\url{https://github.com/shihyuntang/igrins_rv}} developed by our group \citep{stahl2021,tang2021a}. A detailed description of \texttt{IGRINS RV} can be found in \citet{stahl2021}; here, we provide only a brief summary of the package. 

\texttt{IGRINS RV} is designed to exploit the unique advantages of the IGRINS spectrograph. The broad wavelength coverage, high resolution, and mobile, compact design of IGRINS has made it a powerful tool for efforts such as characterizing ultra-cool objects' atmospheric compositions \citep{mansfield2022,tannock2022}, mapping young stellar objects' mean surface magnetic fields \citep{sokal2020} and fundamental parameters \citep{lopez-valdivia2021}, as well as constraining the mass of exoplanets \citep{mann2022}. At the same time, IGRINS was not designed with sub-\ms{} RV precision in mind. The spectrograph is not pressure and temperature controlled. Without a built-in gas cell, laser frequency comb, or Fabry–Perot etalon, determining precision RVs from IGRINS data presents a challenge. \texttt{IGRINS RV} was designed to overcome this obstacle by using a modified forward modeling technique that takes advantage of the Earth's atmospheric absorption lines (telluric lines) as a wavelength calibrator. The code uses the Telfit \citep{gullikson2014} package to create synthetic telluric templates based on fits to telluric standard star observations taken shortly before or after science targets. This produces telluric templates that are high-resolution yet sensitive to the variability of telluric absorption. Stellar templates are generated with the SYNTHMAG C++ code \citep{kochukhov2007} using the VALD line database \citep{ryabchikova2015} and Phoenix ``next generation'' (NextGen) model atmospheres \citep{allard1997,hauschildt1999}.

The accuracy and precision delivered by \texttt{IGRINS RV v1.0} \citep{tang2021a} was validated through long-term monitoring of two RV standard stars (GJ\,281 and HD\,26257) and two known planet-hosting stars ($\tau$~Boo and HD~189733), with estimated RV precisions of $\sim$27 \ms{} in the $K$ band and $\sim$31 \ms{} in the $H$ band typical for slowly rotating targets. However, the code could only measure relative (not absolute) RVs to such precisions. This is because, for a given target, RVs derived from different spectral orders were affected by constant zero-point differences \citep[Figure~6 in][]{stahl2021}. \texttt{IGRINS RV v1.0} simply subtracts out these zero-point offsets to achieve more accurate and precise relative RVs. This approach was shown to be robust based on the success of recovering the planet-induced RV signal from the $\tau$\,Boo and HD~189733 systems \citep{stahl2021}. Nevertheless, users still had the option to run the code in ``absolute RV'' mode, letting the zero-point offsets remain and sacrificing precision.

Subtracting these zero-point offsets, however, is only feasible when they can be accurately estimated. When a target only has a few observations\,---\,such as DI\,Tau, for which we only have five\,---\,then an alternative solution is needed. Here, we introduce \texttt{IGRINS RV v1.5.1}, which almost completely removes the zero-point offset issue in the $K$ band (Figure~\ref{fig:version_copmare}) and includes a more robust treatment of offsets in the rare cases when they remain. Beyond a number of smaller improvements, the biggest change to the pipeline involves the masking of \htwoo{} absorption lines during the spectral fitting process, as these lines were skewing the derived wavelength solutions. More information on \texttt{IGRINS RV v1.5.1} is provided in Appendix~\ref{sec:v1.5.1}. The upgrade to \texttt{v1.5.1} significantly improves accuracy while maintaining the precision of \texttt{v1.0}. The new improvements particularly benefit targets with only a handful of observations ($\lesssim$10). We use \texttt{IGRINS RV v1.5.1} with a synthetic stellar template of \teff{} = 3800 K and \logg{} = 4.0 to measure $K$ band RVs for DI\,Tau, providing the results shown in Table~\ref{tab:rv}. The \teff{} and \logg{} adopted are the products of the magnetic field strength study described in Section~\ref{sec:B}.

\begin{figure}[tb!]
\centering
\includegraphics[width=.9\columnwidth]{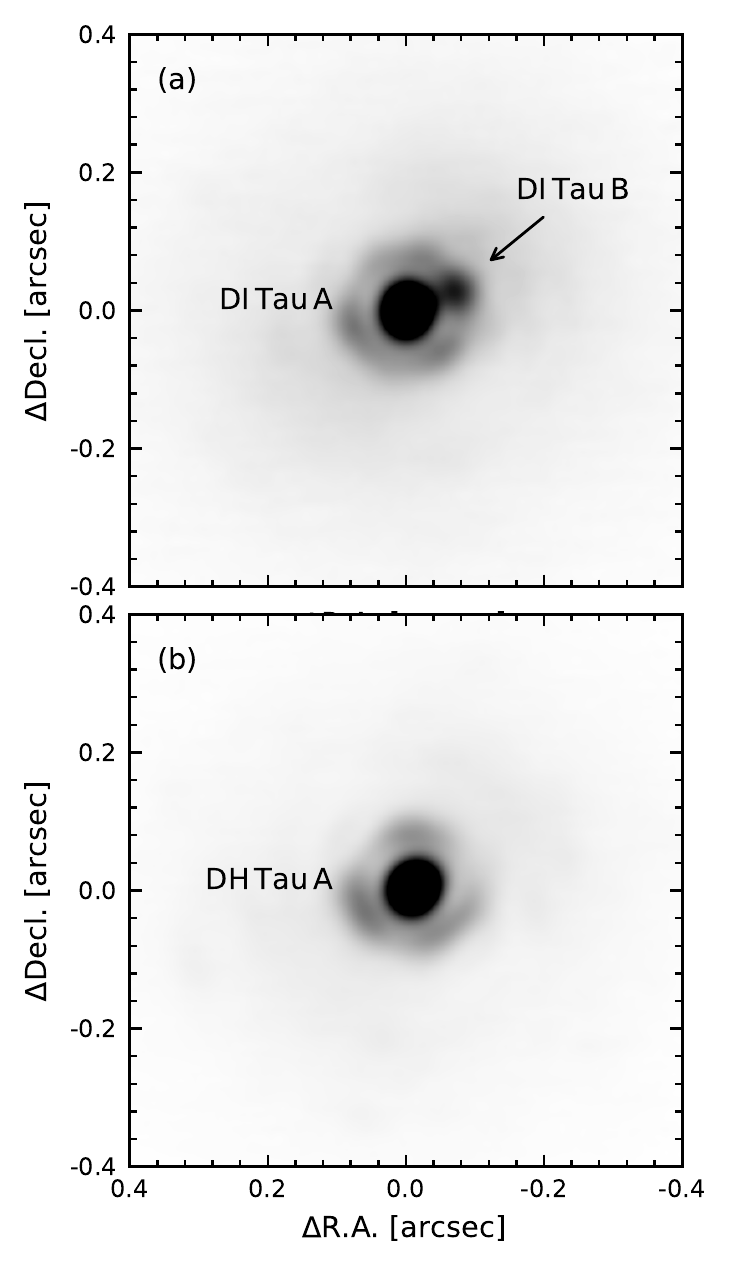}
\caption{
    Keck NIRC2 AO coadded image with K$_{\rm cont}$ filter on UT 2020 February 13. 
    (a) DI\,Tau\,AB, and (b) DH\,Tau\,A, the PSF calibration source.
    DH\,Tau\,B, with a projected separation of 2\farcs3 from DH\,Tau\,A, is outside 
    the plotting region, and is not affecting the PSF subtraction result for DI\,Tau.
    }
\label{fig:AO}
\end{figure}

\subsection{Keck NIRC2 Adaptive Optics Imaging}\label{sec:keckAO}

We obtained AO images for DI\,Tau on UT 2022 February 13, October 19, and December 29 using the near-infrared camera \citep[NIRC2,][]{wizinowich2000} on the 10 m Keck\,II Telescope at the W. M. Keck Observatory. The images were taken using the narrow-field camera that gives a field of view of 10\arcsec{} square. On the first night we collected 18 images in the K$_{\rm cont}$ filter with 3 s exposure times and 1 coadd each, dithered by 3\arcsec{} across the detector. On the other two nights we collected sets of 12 images in four different filters (J$_{\rm cont}$, H$_{\rm cont}$, K$_{\rm cont}$, $L'$) with 0.18--1.0 s exposure times and 10 coadds each, dithered by 2\arcsec{} across the detector. Immediately following the observations of DI\,Tau for each night, we observed a nearby single star (DH\,Tau\,A on 2022 February and DN\,Tau on 2022 October and December) as a point spread function (PSF) reference using the same AO frame rate (1054 Hz on 2022 February and 438 Hz on 2022 October -- December). The images were flatfielded using dark-subtracted dome flats, and the sky background was removed by subtracting pairs of dithered images. 

Figure~\ref{fig:AO} shows the coadded image of DI\,Tau\,AB in panel (a) and DH\,Tau\,A in panel (b) for comparison. Because DI\,Tau\,A and B's Airy rings overlap with each other, we used the separate observations of DH\,Tau and DN\,Tau as a PSF reference to measure the relative separation ($\rho$), position angle (P.A.), and flux ratio ($f_2/f_1$) of DI\,Tau\.B relative to A following the PSF fitting techniques described by \citet{schaefer2006,schaefer2014}. In summary, we constructed a binary model by summing two images of the PSF together and varying the relative position and flux ratio of the components through a grid search procedure to find the combination that gave the minimum $\chi ^2$ value. We then corrected the positions for geometric distortions in the detector, adopted a plate scale of 9.971 $\pm$ 0.004 mas~pixel$^{-1}$, and subtracted an angle of 0\fdg262 $\pm$ 0\fdg020 from the raw P.A. to correct for the orientation of the camera relative to true north \citep{service2016}. The final values and uncertainties for the positions and flux ratios were computed from the mean and standard deviation of results from multiple images. In 2022 October and December, the positions were averaged from the H$_{\rm cont}$ and K$_{\rm cont}$ images as a compromise between the stability of the AO corrected PSF and angular resolution. The final measurements for DI\,Tau are given in Table~\ref{tab:relA}.

\subsection{Lowell $V$ band Photometry}\label{sec:photomery}

CCD photometry for DI\,Tau was obtained between 2012 to 2020 using the Lowell Observatory's 0.7 m telescope in robotic mode. The exposure times were typically 3 minutes through a $V$ band filter. Two or more visits were obtained on every night of observation. The data were reduced via conventional differential aperture photometry using the MPO Canopus software. The approximate $V$ band zero-point was determined using five comparison stars (TYC 1837-0139-1, GSC 1837-0200, GSC 1837-0252, GSC 1837-0318, and GSC 1837-0087). Since 2020, observations have been taken using Lowell Observatory's Hall 1.1 m telescope with a $V$ band filter and 30 s exposures. The comparison stars used for the calibration and reduction procedures were the same as those used with the 0.7 m telescope data. The resulting 871 measurements are given in Table~\ref{tab:photo}. The average nightly internal uncertainly was $\sim$0.007 mag per observation.

\begin{deluxetable}{CCC}
\tablewidth{100pt}
\tablecaption{DI\,Tau Lowell Observatory $V$ Photometry\label{tab:photo}
             }
\tablehead{
    \colhead{\hspace{1.0cm}HJD}\hspace{1.0cm}   &
    \colhead{\hspace{0.8cm}$V$}\hspace{0.8cm}   &
    \colhead{\hspace{0.8cm}$\sigma_V$}\hspace{0.8cm} \\
	\colhead{(days)}    & \colhead{(mag)}   & \colhead{(mag)}
     }
\startdata 
2456256.62833 & 12.916 & 0.003 \\
2456256.63051 & 12.922 & 0.003 \\
2456256.73847 & 12.921 & 0.003 \\
2456256.74065 & 12.920 & 0.002 \\
2456256.85653 & 12.917 & 0.002 \\
2456256.85870 & 12.916 & 0.002 \\
2456256.93987 & 12.914 & 0.002 \\
2456256.94205 & 12.918 & 0.002 \\
2456257.01213 & 12.911 & 0.002 \\
2456257.01430 & 12.911 & 0.002 \\
\hline \\[-9pt]
\enddata
\tablecomments{
    This table is available in its entirety in machine-readable form. Only the first 10 rows are shown here.
    }
\end{deluxetable}

\begin{deluxetable}{l c C}
\tablecaption{DI\,Tau\,AB Preliminary Orbital Parameters\label{tab:par}
             }
\tablehead{
	 \colhead{Parameter}    &  \colhead{Units}  & \colhead{Value}
     }
\startdata 
Period ($P$)                                & yr            & 35.1^{+36.0}_{-2.9}  \\[2pt]
Time of periastron passage ($T_0$)          & JY            & 2017.32^{+0.30}_{-0.23}  \\[2pt]
Eccentricity ($e$)                          &               & 0.734^{+0.128}_{-0.026} \\[2pt]
Semi-major axes ($a$)                       & mas           & 76.8^{+40.3}_{-1.4}    \\[2pt]
Inclination angle ($i$)                     & deg           & 81.2^{+27.6}_{-6.4}  \\[2pt]
Longitude of the ascending node ($\Omega$)  & deg           & 107.9^{+12.4}_{-1.5}    \\[2pt]
Argument of periastron ($\omega_{\rm pri}$) & deg           & 236.2^{+12.9}_{-9.3}   \\[2pt]
Primary RV semi-amplitude ($K_{\rm pri}$)   & \kms          & 1.85^{+0.28}_{-0.18}  \\[2pt]
Opt. system RV ($\gamma_{\rm opt}$)         & \kms          & 15.35^{+0.24}_{-0.12}  \\[2pt]
NIR system RV ($\gamma_{\rm nir}$)          & \kms          & 15.90^{+0.40}_{-0.45}  \\[2pt]
\multicolumn{3}{c}{{\it Derived parameters} } \\
Total mass ($M_{\rm tol}$)                  & $M_{\odot}$   & 0.95^{+0.37}_{-0.35} \\[2pt]
Primary mass ($M_{\rm pri}$)                & $M_{\odot}$   & 0.82 \pm 0.32 \\[2pt]
Secondary mass ($M_{\rm sec}$)              & $M_{\odot}$   & 0.14^{+0.07}_{-0.04} \\
Mass ratio ($M_{\rm sec}$/M$_{\rm pri}$)    &               & 0.17^{+0.07}_{-0.02} \\[2pt]
Semi-major axes ($a$)                       & AU            & 10.6^{+5.5}_{-0.2}   \\[2pt]
\hline \\[-9pt]
\enddata
\tablecomments{
Masses and semi-major axes in AU were derived using the Gaia distance of 137.35 $\pm$ 0.86 pc.
}
\end{deluxetable}

\section{Visual Orbit and RV Joint Fit} \label{sec:orbit_fit}

We used the RVs (Table~\ref{tab:rv}) and the direct imaging data (Table~\ref{tab:relA}) to compute a joint orbital fit and solve for the period ($P$), time of periastron passage ($T_0$), eccentricity ($e$), semi-major axis ($a$), inclination ($i$), longitude of the ascending node ($\Omega$), argument of periastron ($\omega_{\rm pri}$), and RV semi-amplitude of the primary star ($K_{\rm pri}$). To account for possible offsets between the optical and NIR RV measurements, we solved for the system velocity separately for the two sets ($\gamma_{\rm opt}$ and $\gamma_{\rm nir}$, respectively). To investigate the quality of each data set, we initially fit an orbit separately to the optical RVs and the NIR RVs. For a fit to only the NIR RVs, the reduced $\chi^2$ was 6.1, suggesting that the NIR uncertainties are underestimated. We increased the uncertainties for the NIR RVs by a factor of 2.5 compared with those reported in Table~\ref{tab:rv} to force the reduced $\chi^2$ to 1, so that the uncertainties more adequately reflect the scatter in the data.

We explored the range of orbital solutions that fit the data by randomly selecting each of the 10 orbital parameters from a broad range of possible values. For each iteration, we started with the randomly drawn parameters as initial values and optimized the fit by using a Newton-Raphson method to minimize the $\chi^2$ by calculating a first-order Taylor expansion for the equations of orbital motion. We performed 1,000 iterations and selected the orbital solution with the lowest $\chi^2$ value as the best-fit. The final orbital parameters are given in Table~\ref{tab:par}.

The orbital fit is complicated by several factors. The three higher precision Keck AO measurements span a very small portion of the visual orbit ($<$1 yr). They are complimented by two additional observations that span a $\sim$30 yr time frame, but the lunar occultation \citep{chen1990} is only a one-dimensional measurement of the projected separation along the direction of the occultation, and the early speckle observation \citep{ghez1993a} has a larger 10 mas uncertainty in position. The RV measurements fortuitously sampled the most recent periastron passage; however, the RV curve is relatively flat outside of the peak amplitude, and the RV measurements span only about half of a period for the best-fit orbit, so they do not provide a strong constraint on the period.

\begin{figure*}[tb!]
\centering
\includegraphics[width=1.\textwidth]{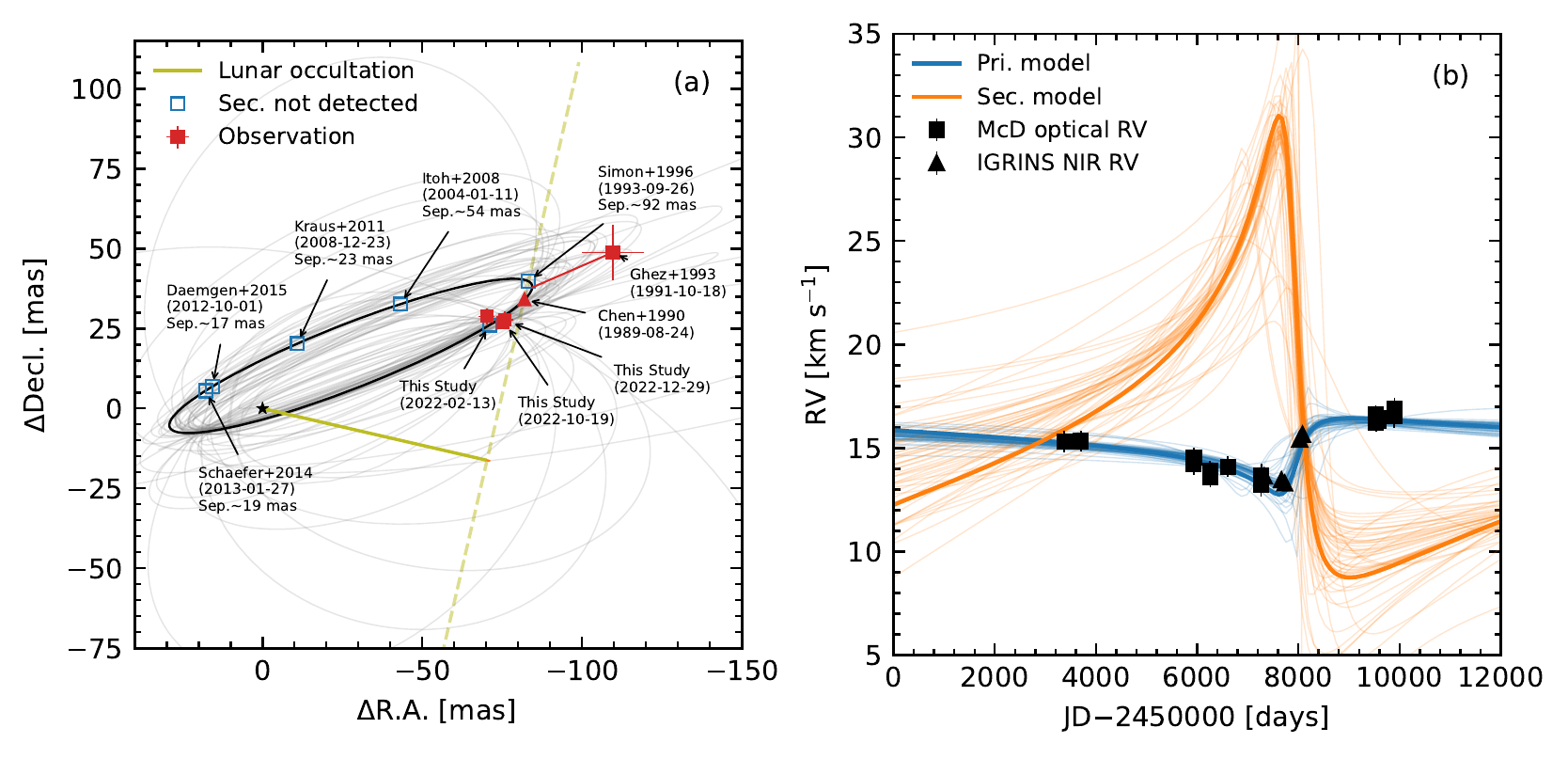}
\caption{
    Model orbits of DI\,Tau\,AB. 
    (a) The black thick line shows the best fit orbit and 
        the grey shaded lines are other possible orbits from the 50 random 
        sampling draw from the bootstrap solutions (Section~\ref{sec:orbit_fit}).
        Red solid squares represent the astrometry from Table~\ref{tab:relA}. 
        The olive color lines are the lunar occultation measurement from \citet{chen1990}, 
        where the dashed line shows the direction of the lunar occultation along a position 
        angle of 257 deg, and the solid line shows the projected separation of 72.1 mas. 
        The red triangle shows the intersection of the model to the lunar occultation measurement. 
        The blue open squares show locations of the secondary based on the best fit orbit model 
        for the observations which did not resolve DI\,Tau\,B (see Section~\ref{sec:non_detections}); the references and separations at the time of observation for the best fit orbit model are indicated.
   (b) Best-fit radial velocity (RV) curve for the primary (thick blue line) and the 
        secondary (thick orange line). The thin shaded lines are the 50 randomly sampled
        bootstrap solutions. The observed RVs from Table~\ref{tab:rv} are shown in black squares for optical RVs and in black triangles for NIR RVs. 
    }
\label{fig:rvorb}
\end{figure*}

To better assess the quality of the orbital fit and determine reasonable uncertainties, we performed a Monte Carlo (MC) bootstrap analysis. We randomly selected measurements, with replacement, from the RV and direct imaging data sets, keeping the total number of measurements in each set the same (i.e., optical RVs, NIR RVs, two-dimensional positions, and one-dimensional separations). We then randomly varied the new selected set of observations within their measured uncertainties assuming a Gaussian distribution. Finally, we optimized the orbital parameters using the values in Table~\ref{tab:par} as the starting points, and optimized the fit to the new set of observables using a Newton-Raphson technique. We generated 10\,000 bootstrap samples and computed uncertainties by calculating the range for each orbital parameter that included 68\% of the values around the median of each distribution. The uncertainty ranges are reported in Table~\ref{tab:par}. The overall best-fit orbit and 50 random samples of the bootstrap solutions are overplotted in Figure~\ref{fig:rvorb}. Corner plots of the parameter distributions are shown in Appendix~\ref{sec:corner_orb} Figure~\ref{fig:corner_orb}.

The dynamical total mass ($M_{\rm tot}$) of the system and the individual masses of the two components of DI\,Tau were then computed using the best-fit orbital parameters, adopting the Gaia distance of 137.35 $\pm$ 0.86 pc. This yielded $M_{\rm tot} = 0.95^{+0.37}_{-0.35} M_{\odot}$, $M_{\rm pri} = 0.82\pm0.32 M_{\odot}$ and $M_{\rm sec} = 0.14^{+0.07}_{-0.04} M_{\odot}$.

\section{Magnetic Field Strength} \label{sec:B}

Strong magnetic ($B$) fields are a hallmark of TTSs, helping produce their characteristic strong high energy (e.g., X-ray) emission and playing an essential role in how CTTSs interact with their disks \citep{bouvier2007,hartmann2016}. Because of the wavelength dependence of the Zeeman effect ($\propto \lambda^2$) compared to that of Doppler broadening ($\propto \lambda^1$), the $K$ band is one of the most sensitive wavelength regimes for measuring stellar magnetic fields \citep{johns-krull2007}. We use IGRINS $K$ band data to measure the magnetic field on DI\,Tau. As DI\,Tau\,A dominates the observed flux, the measured magnetic field is attributed to the primary component only.

\begin{figure*}[tb!]
\centering
\includegraphics[width=.9\textwidth]{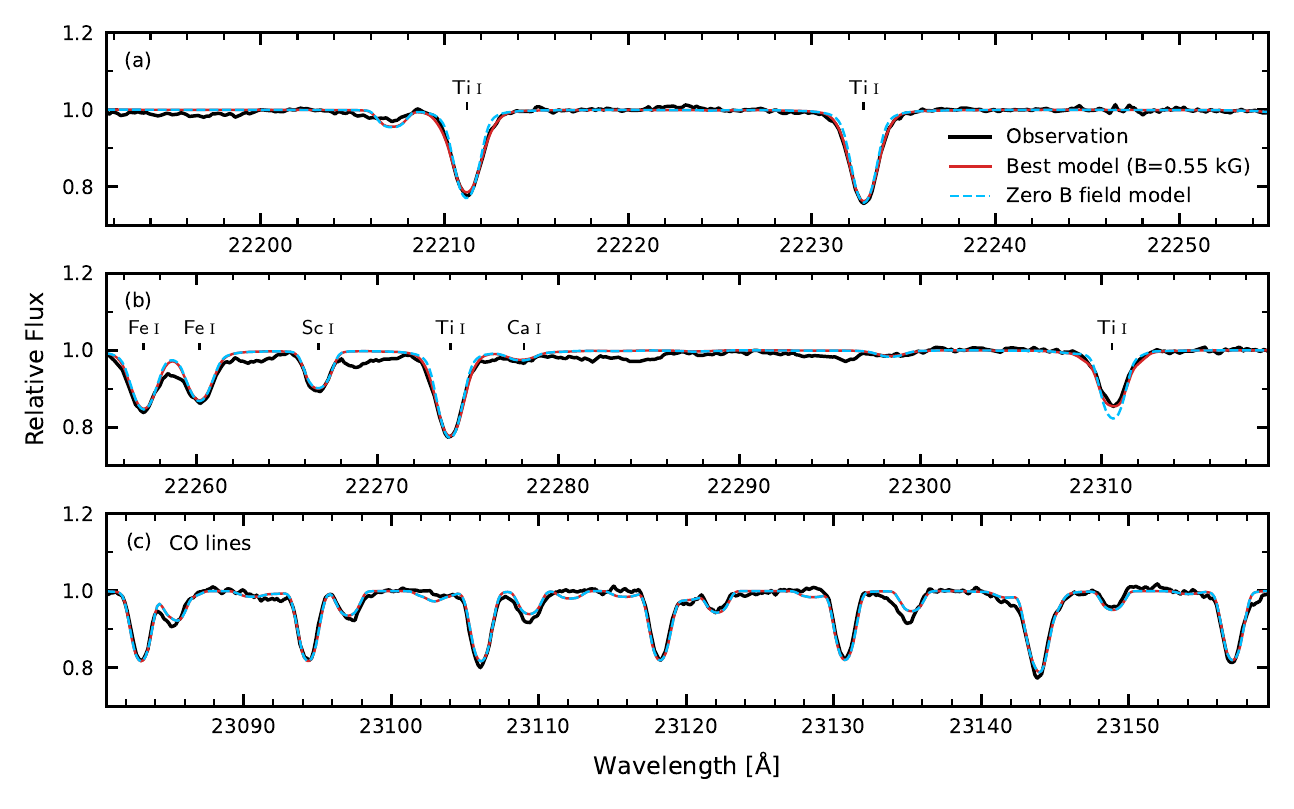}
\caption{
   SYNTHMAG model fits to the observed IGRINS DI\,Tau spectrum for B = 0.55\,kG and B = 0\,kG. Panels (a) and (b) show the magnetic sensitive Ti{\sc i} lines and panel (c) shows the magnetically insensitive CO lines. IGRINS spectra are shown in black lines, and the best fit model with \teff{} = 3800\,K, \logg{} = 4.0, B = 0.55\,kG is shown in red. The zero B field model spectrum is shown in cyan for comparison.
    }
\label{fig:Bfield}
\end{figure*}

We follow the approaches in \citet{johns-krull2007} and \citet{sokal2020}, measuring the magnetic field on DI\,Tau by looking for excess broadening in magnetically sensitive Ti\,{\sc i} lines near 2.225 $\mu$m (Figure~\ref{fig:Bfield} panels a and b). The magnetically insensitive CO lines near 2.312 $\mu$m (Figure~\ref{fig:Bfield}~c) serve as a check on all other non-magnetic line broadening mechanisms that may be present. For a first-order check, we compare the Ti\,{\sc i} lines to the CO lines and found similar line widths, indicating a relatively weak magnetic field. For our detailed analysis, we coadded all five IGRINS $K$ band spectra to obtain the highest possible S/N ($\sim$220 in the continuum) spectrum.  

We use SYNTHMAG \citep{piskunov1999} to compute synthetic spectra. When generating spectra with a magnetic field present, the field is assumed to be radially oriented at the star's surface. A NextGen model atmosphere \citep{allard1995,allard1997,hauschildt1999} was used for computing the synthetic spectra. These model atmospheres are tabulated on a regular grid of effective temperature, surface gravity, and metallicity. \citet{herczeg2014} estimate \teff$= 3774 \pm 88$ K for DI Tau.  For our spectrum synthesis, we assume solar metallicity for DI\,Tau and choose a model from the grid that most closely matches the temperature for DI\,Tau (\teff{} = 3800\,K) and adopt \logg{} = 4.0, which is typical for young stars in Taurus \citep{lopez-valdivia2021}. We adopt a microturbulent broadening of 1 \kms{} and a radial–tangential macroturbulence of 2.0 \kms{} because they are likely appropriate for a star with DI\,Tau’s parameters \citep{gray2005}; our results are very insensitive to these specific values because other line broadening mechanisms (rotation and magnetic) dominate.

In addition to the magnetically sensitive Ti\,{\sc i} lines, the $K$ band contains a number of relatively magnetically insensitive lines, such as the CO lines of the $\nu = 2-0$ rovibrational transitions near 2.3 $\mu$m. Between the two pairs of Ti\,{\sc i} lines are four fairly strong lines of Fe\,{\sc i}, Sc\,{\sc i}, and Ca\,{\sc i} that we include in the spectrum synthesis (Figure~\ref{fig:Bfield}~b). For all lines, we obtained the basic data from the VALD atomic line database \citep{kupka1999} and computed their Zeeman splitting patterns under LS coupling from the transition data contained in the database. In addition, VALD returns several other weak lines in this region which we retain for the spectra synthesis. For the 4 Ti\,{\sc i} lines, the oscillator strengths (log\,$gf$ values) were tuned \citep{johns-krull2004}. This process was not carried out for other lines in this region; we follow the procedure used in other studies \citep[e.g.,][]{johns-krull2004,yang2011} and do not use the magnetically insensitive lines when fitting the synthetic spectra. The role played by these additional lines is to prevent the code from unnecessarily fitting magnetic components to non-magnetic features (e.g., in the far wings of the Ti\,{\sc i} lines) and also help ensure that the fit to the veiling \citep[$r$, non-photospheric excess fluxes,][]{bertout1988} is driven by the depth of the Ti\,{\sc i} lines. 

\subsection{Multi-Component Fit} \label{sec:method1}

We fit three different models for the spectra of DI\,Tau to estimate the magnetic field on the star. In all three fits, we determined the $K$ band veiling ($r_K$) in the Ti\,{\sc i} and CO spectral regions separately. The best way to fit for the magnetic field using the $K$ band spectra is to allow for a distribution of magnetic field strengths on the stellar surface \citep{sokal2020}. Because of the finite spectral resolution and the intrinsic width of the photospheric line profiles (caused by, e.g., thermal, turbulent, and rotational broadening), we only allow for a limited number of magnetic field components when fitting the observed spectra. Our past studies \citep[e.g.,][]{johns-krull2004,johns-krull2007} found that a 2 kG resolution in the field results in fairly robust fits. A significantly finer resolution in the allowed magnetic field strengths will only cause fitted distributions to oscillate substantially. As a result, we fit model spectra for field strengths of 0, 2, 4, and 6 kG, solving for the filling factor of each of these field components. 

We set the resolution of the model spectra to $R = 60\,000$, matching our IGRINS observations in the wavelength regions of interest. We then performed a first set of fits; allowing the stellar \vsini{} to float and found a best-fit value of 12.7 \kms\footnote{
    The \vsini{} of 12.7 \kms{} is slightly higher then the 12.5 \kms{} determined by \texttt{IGRINS RV}; however, this small different results in a different $\bar B$ well within 1$\sigma$ of the final result. 
}. We use this value in the subsequent fits to the magnetic field. The final multi-component fit resulted in filling factors of $f(0$ kG) = 0.75, $f(2$ kG) = 0.23, $f(4$ kG) = 0.00, and $f(6$ kG) = 0.02, which gives a mean field of $\bar B = \Sigma Bf = 0.56$ kG. The best-fit veiling values were 0.14 in the Ti\,{\sc i} region and 0.23 in the CO region. By fixing the \vsini{}, the only free component in the CO region is the veiling.

\subsection{One Magnetic Component and a Single Zero Field Component Fit} \label{sec:method2}

TTSs typically show significantly stronger fields in the 2--3 kG range \citep[e.g.,][]{johns-krull2007,sokal2020,flores2022} making the above estimated $\bar B$ on DI\,Tau\,A lower than typical. The multi-component magnetic field fit is dominated by the 2 kG component ($f = 0.23$). The very small 6 kG component ($f = 0.02$) may be the result of fitting weak features in the wings that are not actually part of the Ti\,{\sc i} lines (e.g., blends). We therefore tried another model with only one magnetic component and a zero field component. This time, we let the field strength of the magnetic component and its filling factor vary along with the veiling in the two spectral regions. The resulting best fit model has a field strength of 1.8 kG with a filling factor of $f = 0.3$, resulting in a mean magnetic field of $\bar B = 0.54$ kG. The veiling values are 0.12 and 0.23 in the Ti\,{\sc i} and CO regions, respectively.

\subsection{Single Component Fit} \label{sec:method3}

Finally, we performed a third fit in which we assumed that a single mean magnetic field value covers the entire star. This model is motivated by a number of recent studies of TTS \citep[e.g.,][]{sokal2018,sokal2020,flores2019,flores2022} which also (or solely) use the spectrum synthesis code MoogStokes \citep{deen2013} to measure magnetic fields on TTSs. 
Our treatment with this fit is not identical to these other MoogStokes-based studies, though, since such studies typically use MARCS stellar atmospheres \citep{gustafsson2008} instead of NextGen models.

With only a single field value covering the entire star, the best-fit model we find has $\bar B = 0.96$ kG with veiling of 0.29 and 0.23 in the Ti\,{\sc i} and CO spectral regions, respectively. The mean value of this single field component model is significantly stronger than that of the other two models; however, the resulting $\chi^2$ for this model is a factor of 1.65 worse than the two multi-component models, which have nearly identical $\chi^2$ values. For comparison, we also performed a field free fit to the spectra where the two veiling values are the only free parameters. The resulting veiling is 0.11 and 0.23 in the Ti\,{\sc i} and CO regions, respectively, but the $\chi^2$ of this model is a factor of 5.45 worse than the multi-component field models.  

\subsection{Summary of Magnetic Field Strength Analysis} \label{sec:Bsummary}

The uncertainty in the mean field was estimated by MC simulations of the data using the measured S/N in the observed spectra and fitting each data realization the same way we fit the observed spectrum. The standard deviation of these MC fits is 0.031 kG, which we take as the random uncertainty. \citet{yang2005} performed tests of magnetic fits to see how sensitive these were to typical errors in the assumed \teff{} ($\pm 200$ K) and \logg{} ($\pm 0.5$), and concluded that the resulting systematic errors are about 10\% of the field measurement, 0.055 kG in our case. Adding this in quadrature to the random uncertainty, we estimate a mean field uncertainty of 0.063 kG. Although the small difference, 0.02 kG, in the final magnetic field estimates between the two multi-component models is reassuring, we are not fully confident that the 6 kG component ($f = 0.02$) in the multi-component model, which contributes 0.12 kG to that final mean field estimate, is valid, as noted in Section \ref{sec:method2}. Therefore, to be conservative, we assign an uncertainty of 0.10 kG to our final mean magnetic field estimate. We conclude that the mean field on DI\,Tau\,A is $0.55 \pm 0.10$ kG (Figure~\ref{fig:Bfield}) with $r_K \sim$0.20 $\pm$ 0.06 (mean and standard deviation of the veiling measurements from the aforementioned three methods).

\begin{deluxetable}{l c l}
\tablecaption{DI\,Tau Physical Parameters\label{tab:phy_par}
             }
\tablehead{
	 \colhead{Parameter} & \colhead{Units} & \colhead{Value}
     }
\startdata 
\multicolumn{3}{c}{IGIRNS $K$ band Magnetic Fitting} 
\\\hline
Effective temperature ($T_{\rm eff}$)           & K     & 3800 (fix)    \\
Surface gravity (\logg)                         & cgs   & 4.0 (fix)     \\
Projected rotational velocity  (\vsini)         & \kms{}& 12.7          \\
$K$ bandVeiling ($r_{\rm K}$)                 &       & 0.20 $\pm$ 0.06          \\
Mean photosphere magnetic strength ($\bar B$)   & kG    & 0.55 $\pm$ 0.10
\\\hline
\multicolumn{3}{c}{SED Fitting} 
\\\hline
Pri. effective temperature ($T_{\rm eff, pri}$)  & K         & 3900 $^{+16}_{-73}$  \\[2pt]
Sec. effective temperature ($T_{\rm eff, sec}$)  & K         & 2868 $^{+228}_{-283}$ \\[2pt]
Radius ratio (R ratio)                           &           & 0.65 $^{+0.10}_{-0.08}$ \\[2pt]
Reddening ($A_v$)                                & mag       & 0.63 $^{+0.03}_{-0.13}$ \\[2pt]
Log pri. flux ratio ($\log{f_{\rm pri}}$)        &           & $-19.12^{+0.024}_{-0.019}$ \\[2pt]
\multicolumn{3}{c}{{\it Derived parameters} } \\
Pri. luminosity ($L_{\rm pri}$)                  & $L_\odot$ & 0.47 $^{+0.02}_{-0.03}$ \\[2pt]
Sec. luminosity ($L_{\rm sec}$)                  & $L_\odot$ & 0.06 $^{+0.12}_{-0.22}$ \\[2pt]
Pri. radius ($R_{\rm pri}$)                      & $R_\odot$ & 1.68 $\pm$ 0.04 \\[2pt]
Sec. radius ($R_{\rm sec}$)                      & $R_\odot$ & 1.11 $\pm$ 0.15
\\\hline
\multicolumn{3}{c}{Lightcurve} 
\\\hline
Rotation period ($P_{\rm rot}$)                  & day   & 7.709 \\[2pt]
\hline \\[-9pt]
\enddata
\end{deluxetable}

\section{Photometry} \label{sec:photo}

Compared to the pronounced photometric variability in CTTS induced by interactions with the surrounding disk \citep[$\Delta V$ can be up to $\sim$2 or 3 mag,][]{grankin2007}, photometric variations in WTTS are smaller ($\Delta V\sim$0.1 mag) as they are dominated by the effect of spots on the surface of a rotating star \citep{herbst1994,grankin2008}. In extreme cases, variations of $\sim$0.5 mag in the optical flux from WTTS can occur \citep[e.g., V410~Tau and V836~Tau,][]{grankin2008}. To investigate the stability of the unresolved photometry used in the two-component SED fit, we first study the lightcurve of DI\,Tau\,AB system, and then introduce the two-component SED fit in the following section.

\subsection{Photometric Variability} \label{sec:photo_var}

\begin{figure}[tb!]
\centering
\includegraphics[width=1.\columnwidth]{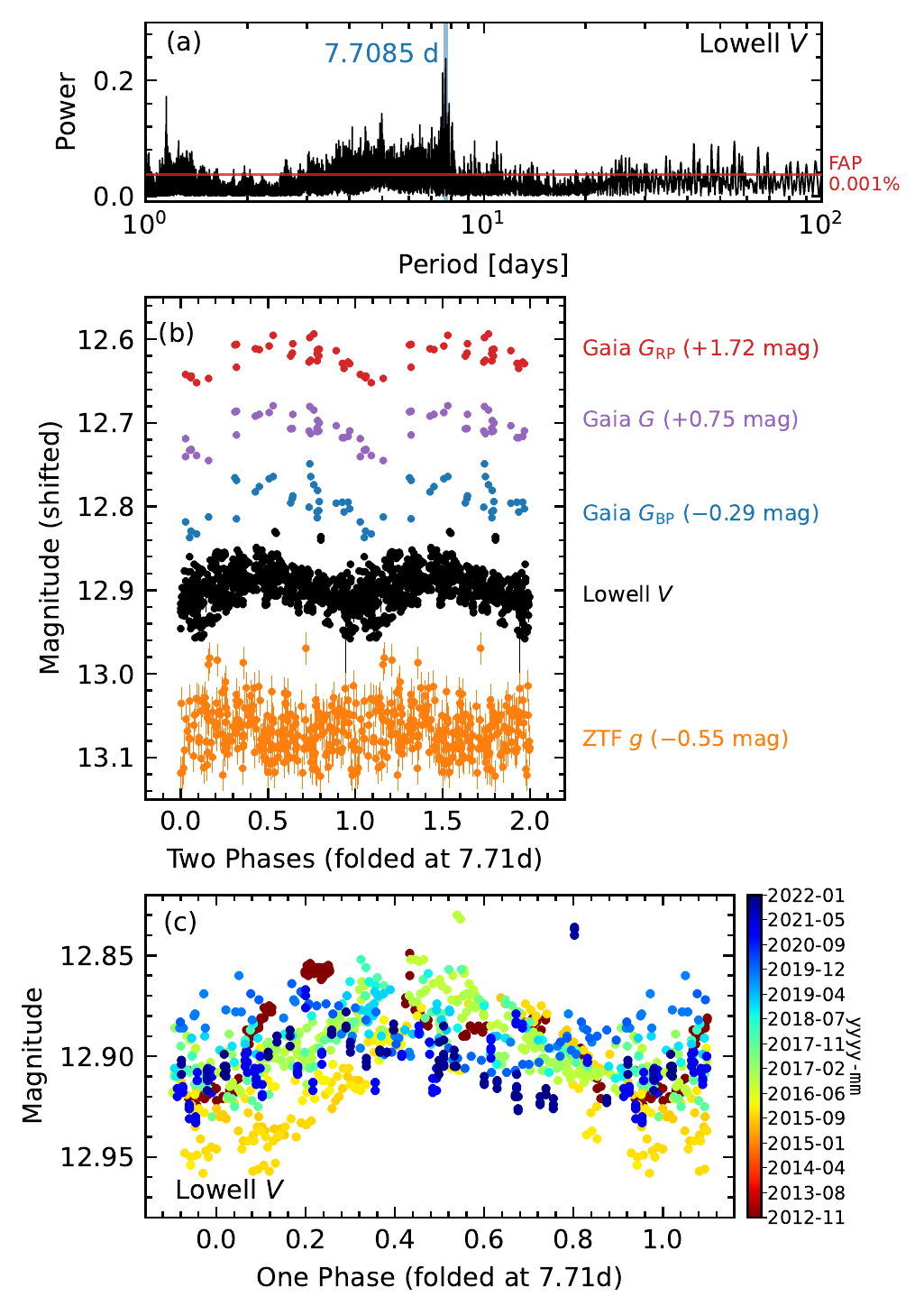}
\caption{
    Results of photometric analysis. (a) Periodogram of the Lowell Observatory $V$ band photometry. The strongest signal at $\sim$7.7085 days is highlighted and the false-alarm probability (FAP) with a power of $\sim$0.05 is shown as the red line. (b) Lightcurves from Gaia $G_{\rm BP}$, $G$, $G_{\rm RP}$, Lowell $V$, and ZTF $g$ folded at the rotational period of 7.71 days. Photometry other than Lowell $V$ are shifted in magnitude for comparison purpose. 
    (c) Lowell $V$ band folded lightcurve color coded by the year and month of observation (color bar). 
    }
\label{fig:photo}
\end{figure}

Figure~\ref{fig:photo}~(a) shows the result of the Lomb-Scargle periodogram analysis for ten years of $V$ band photometric data taken with the Lowell 0.7 m and 1.1 m telescopes (Table~\ref{tab:photo}). The strongest signal in the periodogram at $\sim$7.71 days is similar to results in the literature: e.g., 7.9 days, 7.50 days, 7.733 days \citep[][respectively]{vrba1989,bouvier1993,xiao2012}. The lightcurves from the Lowell $V$ band, Gaia DR3 \citep{gaiacollaboration2022,eyer2022}, and Zwicky Transient Facility \citep[ZTF,][]{bellm2019} data are shown in Figure~\ref{fig:photo}~(b), folded to a period of 7.71 days. In general, similar periodic modulation can be seen in all lightcurves. Yearly changes in DI\,Tau\,AB's $V$ band lightcurve shape can be seen in Figure~\ref{fig:photo}~(c), color-coded by year and month. These shape variations occur because of the evolution\,---\,in terms of size and location\,---\,of spot(s) on the stellar surface. The most active season present in the Lowell $V$ band lightcurve was from 2012 to 2013, showing an amplitude of $\sim$0.06 mag. The quietest season was from 2020 to 2021, with an amplitude of $\sim$0.02 mag. Compared to the $V$ band amplitude of $\sim$0.02 mag from \citet{vrba1989} and $\sim$0.1 mag from \citet{bouvier1993}, we can conclude that DI\,Tau\,AB is relatively inactive, consistent with the relatively long rotation period and the low magnetic field strength (Section \ref{sec:Bsummary}).

\subsection{Two-component SED Fit} \label{sec:sed_fit}

\begin{figure}[tb!]
\centering
\includegraphics[width=1.\columnwidth]{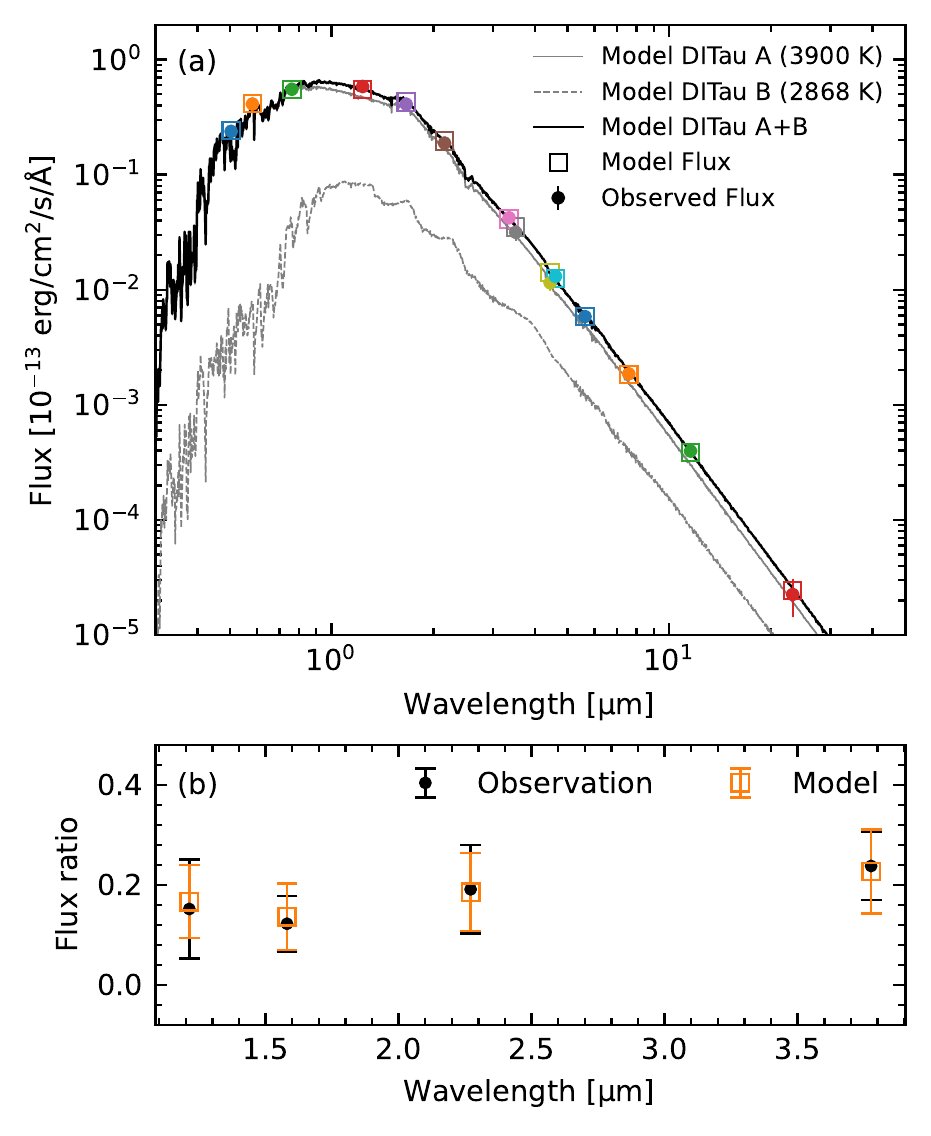}
\caption{
    Two-component SED fit to the unresolved (a) and the resolved (b) photometry of DI\,Tau\,AB. 
    (a) Solid circles are the observed fluxes in Table~\ref{tab:photo}. Band passes for these data, 
        from the shortest to longest wavelengths, are 
        Gaia $G_{\rm BP}$, Gaia $G$, Gaia $G_{\rm RP}$, 2MASS $J$, 2MASS $H$, 2MASS $Ks$, WISE\,W1, 
        IRAC\,36, IRAC\,45, WISE\,W2, IRAC\,58, IRAC\,80, and WISE\,W3. 
        The solid grey line is the best-fit BT-Settl model for DI\,Tau\,A, and the grey dashed line is 
        the best-fit model for DI\,Tau\,B. The final model spectrum for DI\,Tau\,AB is shown with a black 
        line with the integrated flux for each filter shown with open square symbols.
    (b) Keck\,II NIRC2 AO flux ratios for the component-resolved imaging from Table~\ref{tab:relA} are shown 
        with black solid circles and the model-fit results are shown in orange open squares. Band passes for these data, from the shortest to longest wavelengths, are J$_{\rm cont}$, H$_{\rm cont}$, K$_{\rm cont}$, and $L'$.
        The uncertainties for both the unresolved and resolved photometry are 2.9 times larger than 
        those given in Tables~\ref{tab:basic} and \ref{tab:relA} is because of the inflated errors to force a reduced $\chi^2$ of 1 (see section~\ref{sec:sed_fit}).
    }
\label{fig:sed}
\end{figure}

To estimate the individual luminosities and \teff{} for DI\,Tau\,A and B, we used the unresolved (Table~\ref{tab:basic} and Figure~\ref{fig:sed}~a) and resolved photometry (Table~\ref{tab:relA} and shown in Figure~\ref{fig:sed}~b) to fit a two-component model. The thirteen photometric measurements in Table~\ref{tab:basic} were carefully chosen to be uncontaminated by flux from DH\,Tau and to have well-defined filter transmission functions \citep{rodrigo2020}. We did not use the flux ratio from \citet{ghez1993a} because of a lack of proper reference to the accurate filter transmission profile. For the model spectra, we used the \texttt{pysynphot} \citep{stscidevelopmentteam2013} python package to query and interpolate BT-Settl atmosphere model grids \citep{allard2011}, and use the \texttt{pyphot}\footnote{\texttt{pyphot}: \url{https://pypi.org/project/pyphot/}.} python package to integrate the flux within different filters \citep{rodrigo2020}\footnote{SVO Filter Profile Service: \url{http://svo2.cab.inta-csic.es/theory/fps/}.}. For all model spectra used in the SED fitting, we adopted a solar metallicity and \logg{} = 4.0.

The two-component SED fit was done using $\chi ^2$ minimization and adopted the \texttt{NLOpt} python package \citep{johnson2008} for optimization with the Nelder-Mead algorithm \citep{box1965,nelder1965}. The $\chi ^2$ for the unresolved photometry ($\chi^2_{\rm SED}$) is:
\begin{equation}
    \chi^2_{\rm SED} = \sum^{n}_{i=1} \left[ \frac{\left( F_{\rm i,o} - f F_{\rm i,m} \right)^2}{\sigma^2_{\rm i,o}} \right],
\end{equation}
where $n$ is the number of the photometry points, $F_{\rm i,o}$ is the observed flux, $F_{\rm i,m}$ is the total model flux from the primary and the secondary, $f$ is the overall flux scale, and $\sigma^2_{\rm i,o}$ is the error in the observed flux. The overall flux scale is defined as $f \equiv (R/D)^2$, where $R$ is the radius of the stellar object and $D$ is the distance of the object from the Sun, which we fix at 137 pc. The $f F_{\rm i,m}$ can be further expressed as:
\begin{equation}
    f F_{\rm i,m} = f_{\rm pri} F_{\rm i,m,pri} + f_{\rm sec} F_{\rm i,m,sec}.
\end{equation}
The visual extinction ($A_V$) also serves as a free parameter, and is applied to the model spectra based on the reddening law from \citet{cardelli1989} with R(V) = 3.1. 

The $\chi ^2$ for the flux ratio from the resolved photometry ($\chi^2_{\rm fratio}$) has the form:
\begin{equation}
  \chi^2_{\rm fratio} =\sum^{l}_{i=1} \left[ \frac{\left( R_{\rm flux,i,o} - R_{\rm flux,i,m} \right)^2}{\sigma_{\rm i,o}^2} \right],
\end{equation}
where $l$ is the number of the filters in Table~\ref{tab:relA}, $R_{\rm flux,i,o}$ is the observed flux ratio, $R_{\rm flux,i,m}$ is the model flux ratio, and $\sigma^2_{\rm i,o}$ here is the error of observed flux ratio. In total, we fit for five parameters: $T_{\rm eff, pri}$, $T_{\rm eff, sec}$, stellar radius ratio ($R_{\rm sec}/R_{\rm pri}$), $A_V$, and $f_{pri}$. 

The best-fit parameters along with their associated uncertainties (estimated from MC simulations with the data uncertainty increased by 2.9 forcing the $\chi ^2 _\nu$ to 1) are given in Table~\ref{tab:phy_par}. The resulting $A_V$ of 0.63 $^{+0.03}_{-0.13}$ mag is reasonable for a typical WTTS in the Taurus system, and agrees with the value estimated by \citet{herczeg2014} of 0.70$\,\pm\,0.2$ mag. The best-fit model spectra for DI\,Tau\,A and B are shown in Figure~\ref{fig:sed} panel (a) as the grey solid and grey dashed lines, respectively. Figure~\ref{fig:sed} panel (b) shows model fit results of the resolved photometry.

\section{Discussion} \label{sec:discussion}

\subsection{Previous Non-detections of DITau B} \label{sec:non_detections}

In Figure~\ref{fig:rvorb}~(a), the best-fit model's predicted locations for DI\,Tau\,B at the times of observations that resulted in non-detections (Section~\ref{sec:intro}) are highlighted with open blue squares. On UT 2004 January 11, the separation of the primary and secondary was in the range $\sim$40 to $\sim$83 mas, given the uncertainties from the MC bootstrap analysis described in section~\ref{sec:orbit_fit}. With the Subaru CIAO angular resolution of $\sim$100--200 mas FWHM \citep{itoh2008}, DI\,Tau\,B was unresolved. For the observation on UT 2012 October 1, our model predicts a separation range of $\sim$13 to $\sim$52 mas\,---\, too small for Gemini North's NIRI to resolve given the FWHM of the measured PSF, $\sim$80 mas \citep{daemgen2015}. The observation by \citet{schaefer2014} on UT 2013 January 27 occurred near periastron, when our model predicts a separation range of $\sim$14 to $\sim$51 mas, with the best fit model predicted separation of $\sim$19 mas. These predicted separations are below the diffraction limit of $\sim$54 mas for the 10 m Keck telescope in the $K$ band, and are therefore consistent with the non-detection of the companion reported during this epoch.

For the non-redundant masking (NRM) observation taken by \citet{kraus2011} on UT 2008 December 23, our MC bootstrap analysis predicts a separation range of $\sim$15 to $\sim$63 mas; the best-fit model gives a separation of $\sim$23 mas. The detection limit ($\Delta K'$) of DI\,Tau\,B estimated by \citet[][Table~4]{kraus2011} for separations of 10--80 mas is 2.66--5.7 mag, which is larger than our observed $\Delta K$ of 1.47--2.22 mag (Table~\ref{tab:relA}). It is not clear what the cause of the non-detection by \citet{kraus2011} was, but at the minimum separation our orbital solution allows for, the published detection limit is quite close to our maximum measured flux ratio. Thus, if our orbital uncertainties are marginally underestimated, the true separation during December 2008 may actually have been below the NRM detection limit. Similarly, the measurement uncertainty in the NRM results might be marginally underestimated and produced smaller $\Delta K'$ detection limits at the closest separations. Additional observations that improve the precision of the calculated visual orbit could resolve this ambiguity.

The non-detections of DI\,Tau\,B on UT 1993 September 26 and 1993 October 25 were likely the result of DI\,Tau\,B's low flux compared to the detection limit of HST's FGS \citep{simon1996}.  \citet{simon1996} estimated a $V$ band magnitude $\geq$16.5 mag for DI\,Tau\,B, based on their non-detection. Using the best fit model for DI\,Tau\,B from the two-component SED fit in Section ~\ref{sec:evo_model}, a distance of 137 pc, a Johnson $V$ band filter profile \citep{johnson1953}, and the associated Vega zero point, we found a $V$ band magnitude for DI\,Tau\,B of $\sim$18.5 mag, in agreement with the results of \citet{simon1996}. 

\begin{figure}[tb!]
\centering
\includegraphics[width=1.\columnwidth]{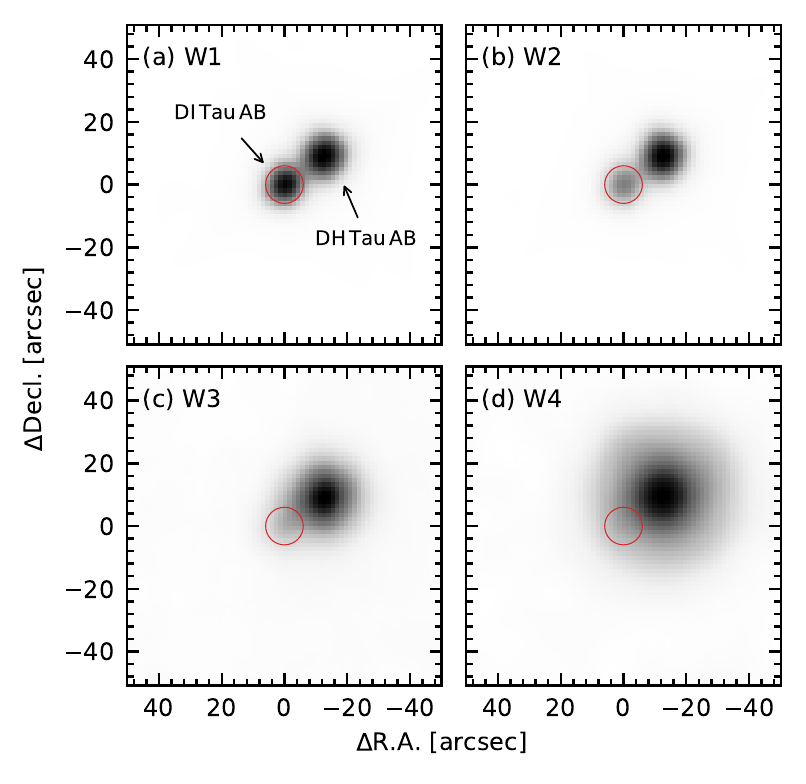}
\caption{
    WISE W1, W2, W3, and W4 images showing DI\,Tau\,AB and DH\,Tau\,AB. Compared to DH\,Tau\,AB, flux from DI\,Tau\,AB falls off rapidly at wavelengths longer than W1.
    The ALLWSIE \citep{cutri2021} W4 magnitude for DI\,Tau\,AB (4.963 $\pm$ 0.032 mag) is contaminated by DH\,Tau\,AB ($\sim$3.032 $\pm$ 0.023 mag). We conclude that no mid-IR (W4) excess is present in DI\,Tau\,AB.}
\label{fig:wise}
\end{figure}

\subsection{Is DITau A a CTTS, a WTTS, or Something in Between?} \label{sec:cttORwtt}

\begin{figure}[tb!]
\centering
\includegraphics[width=1.\columnwidth]{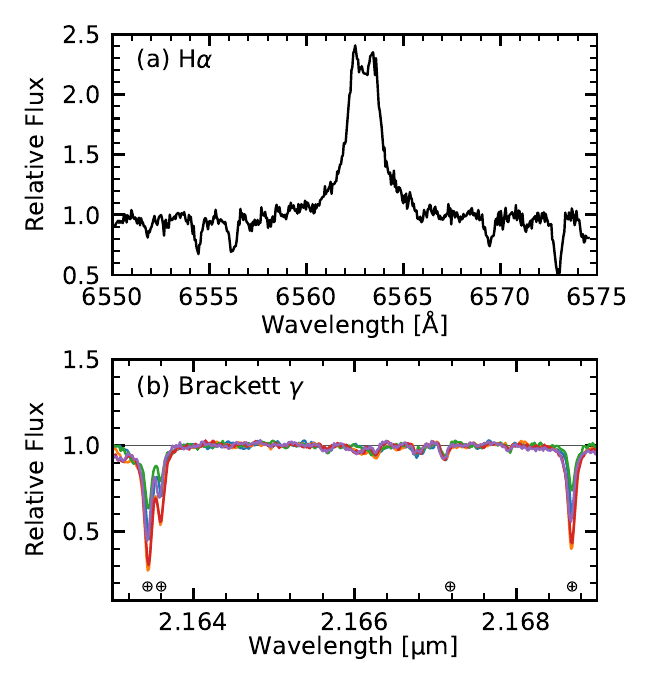}
\caption{Optical and NIR spectra of DI\,Tau around the H$\alpha$ 
            and Brackett $\gamma$ emission lines.
    (a) McDonald optical spectrum from UT 2021 November 13 where
        the H$\alpha$ is most prominent in our data 
        (EW$_\lambda$(H$\alpha$) = 2.9 $\pm$ 0.02 {\r A}).
        The self-absorption around the peak of the H$\alpha$ emission 
        is a non-LTE effect for slowly rotating convective stars \citep{hawley1994}.
    (b) All five epoch of IGRINS K band spectra (color lines).
        The grey horizontal line show the flux = 1 continuum. 
        No emission features can be seen for the Brackett $\gamma$ 
        ($\sim$2.1661 $\rm \mu$m). IGRINS spectra shown here are not 
        been divided by telluric standard star. Four major telluric lines 
        are highlighted with $\oplus$ symbols.
        }
\label{fig:spectra}
\end{figure}

The zero veiling measurements reported in the literature at optical wavelengths \citep[e.g., $r_{7510}$ = 0.0;][]{herczeg2014}, and the absence of Brackett $\gamma$ emission lines and weak H$\alpha$ emission (EW$_\lambda$(H$\alpha$) $\sim$ 1.1 $\pm$ 0.46 {\r A}) in our NIR and optical spectra (Figure~\ref{fig:spectra}), indicate a lack of ongoing accretion at least onto DI\,Tau\,A. Also, the low NIR veiling reported in the literature \citep[e.g., $r_J \sim$0.1 $\pm$ 0.11][]{folha1999}, and found in this study ($r_K \sim$0.2, Section~\ref{sec:B}) indicate that no inner dust disk is present around DI\,Tau\,A. The only evidence supporting a circumstellar disk in the DI\,Tau system was a reported excess of mid-IR flux \citep{kenyon1995,sullivan2022}. Subsequent studies \citep[e.g.,][]{meyer1997,hartmann2005}, however, demonstrated that the IRAS 12 \micron{} excess was contaminated by DH\,Tau and that the WISE W4 excess was contaminated as well, as shown in Figure~\ref{fig:wise}. Therefore, we suggest that DI\,Tau is a WTTS.

\begin{figure}[tb!]
\centering
\includegraphics[width=1.\columnwidth]{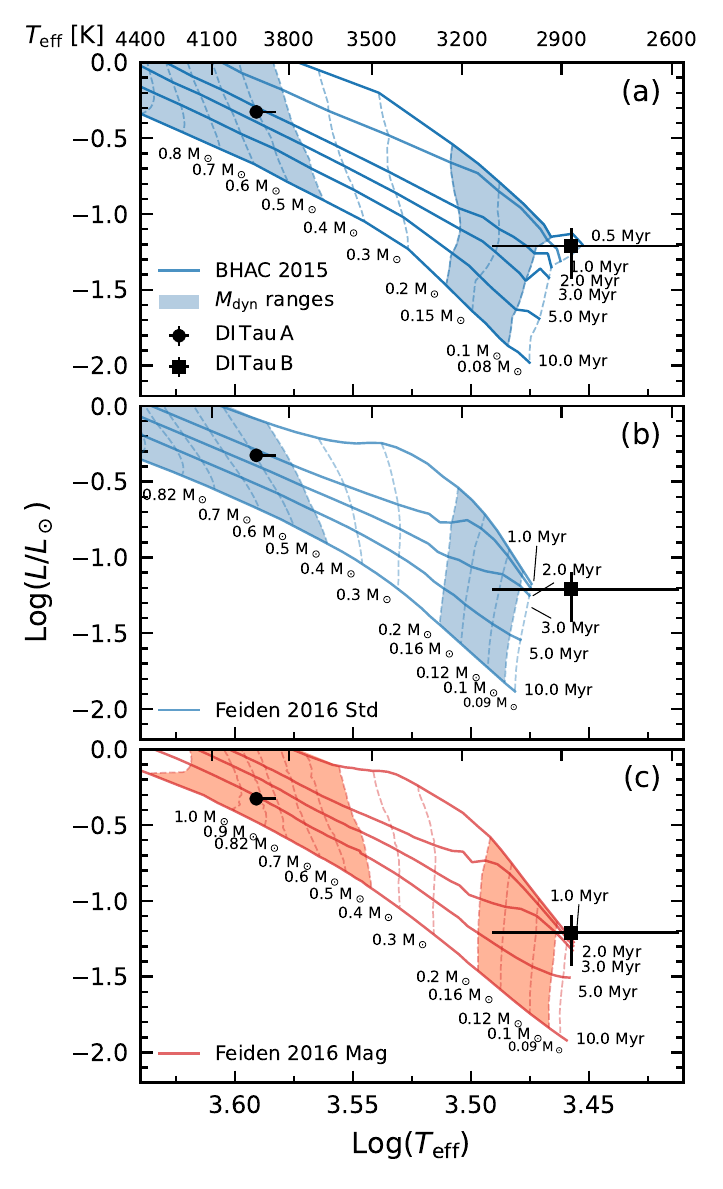}
\caption{
    Hertzsprung-Russell diagram showing DI\,Tau\,A and B with luminosity and effective temperature (\teff{}) from the SED fitting (Section~\ref{sec:sed_fit}). Two standard sets of evolutionary tracks \citep[][BHAC 2015]{baraffe2015} and \citet{feiden2016} are shown in panels (a) and (b), respectively. Panel (c) shows the magnetic version of the evolutionary tracks from \citet[][private communication]{feiden2016}. The colored area in all three panels show the ranges ($\pm1\sigma$) for the dynamical masses, $M_{\rm dyn}$, estimated from the orbital fit in Section~\ref{sec:orbit_fit}.
    }
\label{fig:HRD}
\end{figure}


\subsection{Evolutionary Model Comparison} \label{sec:evo_model}

The Hertzsprung-Russell diagram (HRD) in Figure~\ref{fig:HRD} shows our best-fit results from Section \ref{sec:sed_fit} along with evolutionary models from \citet[][BCAH 2015]{baraffe2015} (panel a) and \citet{feiden2016} (panels b and c). For the Feiden evolutionary models, we show both the standard (Std, panel b) and the magnetic (Mag, private communication, panel c) versions as the latter was shown to be more suitable for young stellar objects \citep[e.g.,][]{feiden2016,simon2019}. 

Both the BHAC 2015 and \citet{feiden2016} Std models indicate that DI\,Tau\,A has an age of $\sim$2--3 Myr and a mass of $\sim$0.62 M$_\odot$, near the lower limit of the dynamical mass, $M_{\rm pri} = 0.82 \pm 0.32 M_{\odot}$ (highlighted as colored blue area in Figure~\ref{fig:HRD}~a and b). We see similar behavior for DI\,Tau\,B between the models' predicted masses and the dynamical mass. Both non-magnetic models predict that DI\,Tau\,B is most likely to have a substellar mass, around 0.08 M$_\odot$, whereas the dynamical mass is $M_{\rm sec} = 0.14^{+0.07}_{-0.04} M_{\odot}$. In comparison to the standard models, the \citet{feiden2016} Mag tracks provide a better fit for the primary, predicting a model mass of about 0.84 M$_\odot$, near the center of the range of possible dynamical masses. However, the uncertainties in \teff{} are large enough that the two components of DI\,Tau\ are consistent with being coeval. Nonetheless, there also remains the possibility that the two stars may have different $\bar B$ field strengths, which would impact their locations on the HRD. 

Identifying any potential differences in the magnetic field strength between DI\,Tau\,A and B requires component-resolved spectroscopy of both stars to directly measure DI\,Tau\,B's $\bar B$. Although challenging, such observations are possible with AO-fed IR spectroscopy, for example, with NIRSPEC on the Keck II telescope \citep[e.g.,][]{prato2002,allen2017}.

\section{Summary} \label{sec:summary}

In this study, we confirm the binary nature of the DI\,Tau system using three epochs of NIRC2 AO imaging, a 17-year span of optical and NIR radial velocity (RV) data, and two historical position measurements from a speckle imaging and lunar occultation observations. These astrometric and spectroscopic data, along with multi-band photometry from sky surveys, allow us to also characterize the physical properties of DI\,Tau\,A and B. Some key highlights and results from this study are summarized as follows:
\begin{itemize}
    \item We model DI\,Tau system's motion and find a best fit orbit that accounts for both historical detections and (almost all of the) non-detections of the secondary.
    \item The small mass ratio, $\sim$0.17, highlights the value of the DI\,Tau system for testing models of PMS evolution as the two stars span a large parameter space across the mass tracks.
    \item With the RV data only covering one periastron passage, and a lack of full orbital sampling from the relative astrometry, our preliminary orbital solution is uncertain and will be revised as additional measurements are made; however, the lower limit on the orbital period, $\sim$32 years, is unlikely to change.
    \item Because of the large uncertainties in the estimated masses from the best-fit orbit resulting from the sparse astrometric and spectroscopic observations, and the large error-bar for the $T_{\rm eff, sec}$ given the low flux ratio between DI\,Tau\,A and B, it is impossible to distinguish between the models shown in Figure~\ref{fig:HRD}. Follow up observations to improve the measurement of the dynamical masses and/or \teff{} are necessary.
    \item We measure an unusually low surface averaged magnetic field strength, $\sim$0.55 kG, for DI\,Tau\,A compared to other TTSs. This may indicate an older age for DI\,Tau\,A \citep{vidotto2014}, although even with the large uncertainties evident in Figure~\ref{fig:HRD}, DI\,Tau\,A is likely younger than 5--6 Myr. Spectroscopic observations of DI\,Tau\,B at adequate signal to noise to establish its magnetic field strength will be important for comparison.
    \item We raise the possibility of non-uniform surface averaged magnetic field strengths in the two components of even relatively short period binaries as a potential source of non-coevality. This can be tested with spectroscopic observations of angularly-resolved close young binaries.
    \item We conclude that DI\,Tau\,A is a WTTS based on the small amplitude $V$ band variability over the last $\sim$40 years, EW$_\lambda$(H$\alpha$) $<$5 {\r A} ($\sim$1.1 $\pm$ 0.46 {\r A}), and a lack of IR excess.
\end{itemize}
\clearpage

\acknowledgments
We are thankful for the comments and suggestions from an anonymous referee to help improving the quality of this paper.
We thank the technical and logistical staff at McDonald and Lowell Observatories for their excellent support of the Immersion Grating Infrared Spectrograph (IGRINS) installations, software, and observation program described here. In particular, D. Doss, C. Gibson, J. Kuehne, K. Meyer, B. Hardesty, F. Cornelius, M. Sweaton, J. Gehring, S. Zoonematkermani, E. Dunham, S. Levine, H. Roe, W. DeGroff, G. Jacoby, T. Pugh, A. Hayslip, and H. Larson. We also thank G. Feiden for providing us his magnetic evolutionary tracks, and A. Kraus, M. Ireland, T. Dupuy, and E. Evans for helpful discussion on the NRM data of DI Tau. Partial support for this work was provided by NASA Exoplanet Research Program grant 80-NSSC19K-0289 (to LP). CMJ acknowledges partial support for this work through grants to Rice University provided by NASA (award 80-NSSC18K-0828) and the NSF (awards AST-2009197 and AST-1461918). GHS acknowledges support from NSF AST-2034336.

We are grateful for the generous donations of John and Ginger Giovale, the BF Foundation, and others which made the IGRINS-LDT program possible. Additional funding for IGRINS at the LDT was provided by the Mt. Cuba Astronomical Foundation and the Orr Family Foundation. IGRINS was developed under a collaboration between the University of Texas at Austin and the Korea Astronomy and Space Science Institute (KASI) with the financial support of the US National Science Foundation under grant AST-1229522 and AST-1702267, of the University of Texas at Austin, and of the Korean GMT Project of KASI.
Some of the data presented herein were obtained at the W. M. Keck Observatory, which is operated as a scientific partnership among the California Institute of Technology, the University of California, and the National Aeronautics and Space Administration. The Observatory was made possible by the generous financial support of the W. M. Keck Foundation. Some of the time at the Keck Observatory was granted by NOIRLab (NOIRLab PropID:2022B-970020; PI: G. Schaefer) through NSF's Mid-Scale Innovations Program. The authors wish to recognize and acknowledge the very significant cultural role and reverence that the summit of Maunakea has always had within the indigenous Hawaiian community.  We are most fortunate to have the opportunity to conduct observations from this mountain.

This work made use of the VALD database, operated at Uppsala University, the Institute of Astronomy RAS in Moscow, and the University of Vienna. This study also made use of the SIMBAD database and the VizieR catalogue access tool, both operated at CDS, Strasbourg, France. Some observations were obtained at the Lowell Discovery Telescope (LDT) at Lowell Observatory. Lowell is a private, non-profit institution dedicated to astrophysical research and public appreciation of astronomy and operates the LDT in partnership with Boston University, the University of Maryland, the University of Toledo, Northern Arizona University and Yale University. We have used IGRINS archival data older than the 2 year proprietary period. This research has also made use of the Spanish Virtual Observatory (\url{https://svo.cab.inta-csic.es}) project funded by MCIN/AEI/10.13039/501100011033/ through grant PID2020-112949GB-I00


\facilities{LDT (IGRINS),
            Smith (IGRINS, Tull Coud\'e Spectrograph),
            Keck:II (NIRC2),
            LO:0.8m, Hall}

\software{
        \texttt{astropy} \citep{astropycollaboration2013,astropycollaboration2018},  
        \texttt{matplotlib} \citep{hunter2007},
        \texttt{NumPy}, \citep{harris2020}
        \texttt{IGRINS RV} \citep{stahl2021,tang2021a},    
        \texttt{PHOEBE} \citep{conroy2020},
        \texttt{pysynphot} \citep{stscidevelopmentteam2013}
        }


\appendix{} 
\counterwithin{figure}{section}
\counterwithin{table}{section}

\section{IGRINS RV v1.5.1}\label{sec:v1.5.1}

The newest version of \texttt{IGRINS RV} provides numerous improvements. A summary of the changes is included below.

\subsection{Zero-point RV Offsets Between Orders}\label{sec:offset}

The largest update is related to the zero-point RV offsets described in Section~\ref{sec:intro}. Further investigation into the origin of these offsets led us to \citet{reiners2016}, who note that the \citet{livingston1991} atlas of the infrared solar spectrum\,---\,which the wavelength calibration of \texttt{IGRINS RV} synthetic telluric templates was based on \citep[see Section 3.3][]{stahl2021}\,---\,itself displays inaccuracies in its wavelength scale of up to hundreds of \ms. This could easily corrupt the accuracy of the synthetic telluric templates generated by \texttt{IGRINS RV}. Without a high-resolution, high-S/N, wavelength-accurate telluric observation spanning the full $H$ and $K$ bands, the telluric template generation process used by \texttt{IGRINS RV} is compromised.

\begin{figure}[tb!]
\centering
\includegraphics[width=1.\columnwidth]{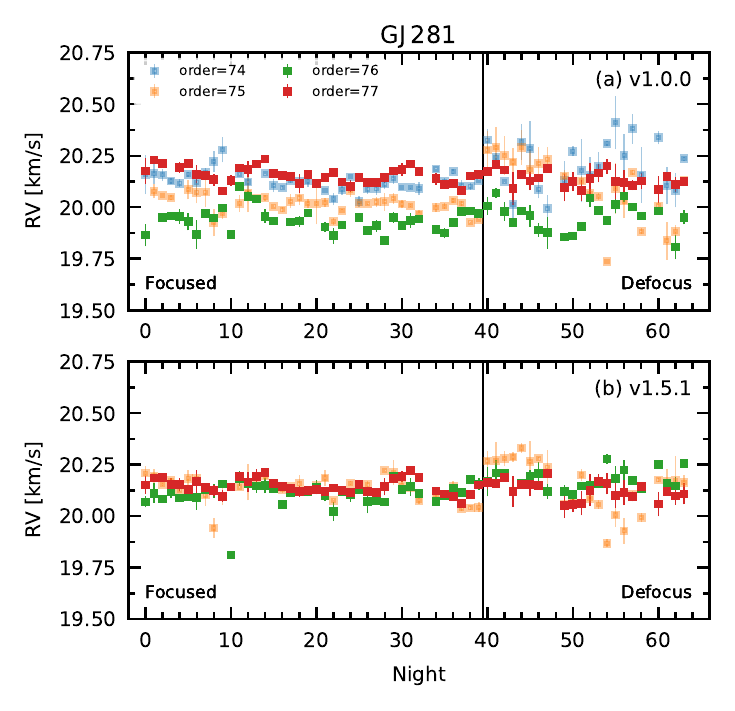}
\caption{
    Data for the RV standard star, GJ\,281. Results (a) from \texttt{IGRINS RV} v1.0.0, and (b) from \texttt{IGRINS RV} v1.5.1. The biggest improvement in v1.5.1 is the correction of zero-point RV offsets between orders in the $K$ band. This upgrade allows \texttt{IGRINS RV} to achieve higher accuracy in absolute RV measurements while retaining the $\sim$30 \ms{} precision.}
\label{fig:version_copmare}
\end{figure}

As noted in \citet{stahl2021}, an alternative would be to rely on the wavelength scale internal to Telfit, which is based on the AER line list (mostly culled from HITRAN 2016). In \citet{stahl2021}, we opted against this method because some HITRAN line data is known to be accurate only to hundreds of \ms{} (or even only on the level of \kms). Yet, a closer inspection of the line data in the wavelength regions of interest showed that it was only the \htwoo{} absorption lines that exhibited such large inaccuracies. If we mask all regions with significant \htwoo{} absorption in our spectral fitting, then using Telfit's internal wavelength scale almost completely eliminates the zero-point RV offsets in the $K$ band (Figure~\ref{fig:version_copmare}~b). This change was incorporated into \texttt{IGRINS RV v1.5.1}, and has the side benefit of significantly reducing the computation time required to generate the synthetic telluric templates.

However, even with this change, some degree of zero-point offsets between results from different orders remained. This is particularly the case when applying \texttt{IGRINS RV} to spectroscopic binaries, in the $H$ band, or to order 75 $K$ band observations taken when the spectrograph suffered from defocus issue \citep[see][Section 2.1]{stahl2021}. After additional testing, we found these offsets to be caused by wavelength-dependent mismatches between our stellar templates and the data. These mismatches occur because: (1) our templates are not perfect\,---\,particularly at lower temperatures, neither PHOENIX \citep{husser2013} nor SYNTHMAG models correspond to our observed spectra as well as they do for targets with \teff{} $\gtrapprox$ 4000\,K, and (2) spots can distort the observed data. This adds an extra absorption component that our applied stellar template is not guaranteed to fit in the same way for every order. The spot-induced absorption is also from a cooler source, which is (as per point 1) harder for our stellar templates to reproduce. We found this explanation to be supported by a variety of experiments that traced the effects of applying different stellar templates to different types of targets, including simulated spectra of spotted stars. 

Relatively warm and unspotted stars exhibit stellar absorption that is broadly well-simulated by our synthetic templates, so the RV offsets between orders for such targets are small. The more heavily spotted or cooler a target is, the less representative our templates become, leading to larger zero-point offsets between orders (Tang et al., in prep.). Spectroscopic binaries also show larger offsets because their cooler secondary stellar components affect spectra similarly to non-polar spots. In the $H$ band, additional investigations are still underway, but it appears that in this waveband the line lists we use to generate our synthetic stellar templates are broadly less accurate. Lastly, the fact that we observe significantly greater RV offsets in order 75 when the spectrograph mounting was loose can be explained by the increased instrumental broadening, which produces lower stellar and telluric information content, which in turn makes for more poorly calibrated wavelength solutions. It is reasonable that this would affect the measurements from order 75 in particular, given that we use the region we use of this order covers the smallest wavelength span and features the fewest absorption lines than the other analysis regions.

\subsection{Updated RV Uncertainly Estimation}\label{sec:unc}

Overall, stellar template mismatch will produce zero-point RV offsets at some level in all targets, so we opted to update our uncertainty calculation to take into account this additional source on a target-by-target basis. The calculation is as follows:

First, for every observation x, the RV and its uncertainty are determined for each order analyzed j, as per Section 3.7 of \citet{stahl2021}. We call these $\rm RV_{xj}$ and $\rm S_{xj}$.

Next, for each order, we compute the weighted mean and the standard deviation of the weighted mean:

\begin{equation}
    \rm w_{x} = \bigg(\frac{1}{S_{xj}^{2}}\bigg)\bigg/\bigg(\sum_{x} \frac{1}{S_{xj}^{2}}\bigg)
\end{equation}

\begin{equation}
    \rm RV_{j} = \sum_{x} (w_{x} \cdot RV_{xj})
\end{equation}

\begin{equation}
    \rm \sigma_{j} = \left(\sqrt{\sum_{x} \frac{1}{S_{xj}^{2}}}\right)^{-1}
\end{equation}

We then check if the weighted means of each order are statistically consistent with each other. For each order k and each order m, where k $\neq$ m, we compare:

\begin{equation} \label{eq:rvorder}
    \rm | RV_{k} - RV_{m} | \stackrel{?}= \left(\sqrt{\sigma_{m}^{2}\ +\ \sigma_{m}^{2}}\right)
\end{equation}
\vspace{3pt}

If the left hand side is less than the right hand side for all values of k and m, then no significant zero-point offsets between orders are detected, and no additional uncertainty calculation is needed. If the left hand side is ever greater than the right hand side, however, we then correct the RVs for each order by the zero-point difference between the average RV of the order and that of the best-performing order q (in the $K$ band, this is order 77). Thus, the new RV for every observation and order becomes:

\begin{equation}
    \rm RV_{xj}' = RV_{xj} + RV_{q} - RV_{j}
\end{equation}
where $\rm RV_{q}$ and $\rm RV_{j}$ are determined by Equation~\ref{eq:rvorder}. We then calculate two additional uncertainty terms. The first takes into account the additional uncertainty that comes from subtracting measurements of the mean RV of each order:

\begin{equation}
    \rm \psi = \sqrt{\sum_{j} \sigma_{j}^{2}}
\end{equation}
\vspace{3pt}

The second uncertainty term is calculated, for each observation, as the amount of scatter observed between the different order RV measurements that cannot be explained by their uncertainties:

\begin{equation}
    \rm \phi_{x} = \sqrt{ \sigma_{x}   - \sum_{j} S_{xj}^{2}}
\end{equation}
where $\sigma_{x}$ is the standard deviation of the mean of $\rm RV_{xj}$ for a given observation x. This characterizes the nightly deviation of the order RVs beyond what we would expect from all the other factors that contribute to our uncertainties. That is, for zero-point offsets that are consistent between observations (because of stable differences between the observed spectra and the stellar template), the mean order correction and the inclusion of $\psi$ ought to suffice\,---\,but in cases when the zero-point offset is variable between observations because of changing spot geometry or secondary stellar components, then $\phi_{x}$ is necessary to describe the additional uncertainty. The final RV of each observation is then computed as per \citet{stahl2021} except with $\rm RV_{xj}'$ instead of $\rm RV_{xj}$, and the corresponding uncertainty is expanded to include $\psi$ and $\phi_x$:

\begin{equation}
    \rm w_{j} = \bigg(\frac{1}{S_{xj}^{2}}\bigg)\bigg/\bigg(\sum_{j} \frac{1}{S_{xj}^{2}}\bigg)
\end{equation}

\begin{equation}
    \rm RV_{x} = \sum_{j} (w_{j} \cdot RV_{xj}')
\end{equation}

The associated uncertainty is:

\begin{equation}
    \rm \sigma_{x}  = \sqrt{ \phi_{x}^{2}  + \psi^{2} + \left(\sum_{j} \frac{1}{S_{xj}^{2}}\right)^{-1}}
\end{equation}
\vspace{3pt}

Typical values of $\psi$ vary between 5--50 \ms, while $\phi_{x}$ is about 0--15 \ms{}, although sometimes can be as high as 70 \ms. Accounting for these new uncertainty measurements, we found that during the period when the spectrograph experienced a defocus, the large zero-point offsets of our lowest-performing order ($\#$75) meant that including it often led to worse precisions than if it were left out. We therefore opt to exclude order 75 from analysis for observations taken when the spectograph was defocussed.

In summary, \texttt{IGRINS RV} v1.5.1 delivers a better accuracy in the absolute RV by trading off in the relative RV precision, $\sim$30.4 \ms{} compared to $\sim$26.8 \ms{} in version v1.0.0's. Lastly, a number of smaller changes were made to the code, including:

\begin{itemize}
    \item Included \texttt{numba} to streamline code speed 
    \item Updated Telfit compiler to avoid critical error that popped up in some cases
    \item Corrected bugs related to over-interpolation of templates
    \item Dropped Order $\#$74 from $K$ band analysis 
\end{itemize}

\section{Corner Plot for the Orbital Fit}\label{sec:corner_orb}
\begin{figure*}[tb!]
\centering
\includegraphics[width=1.\textwidth]{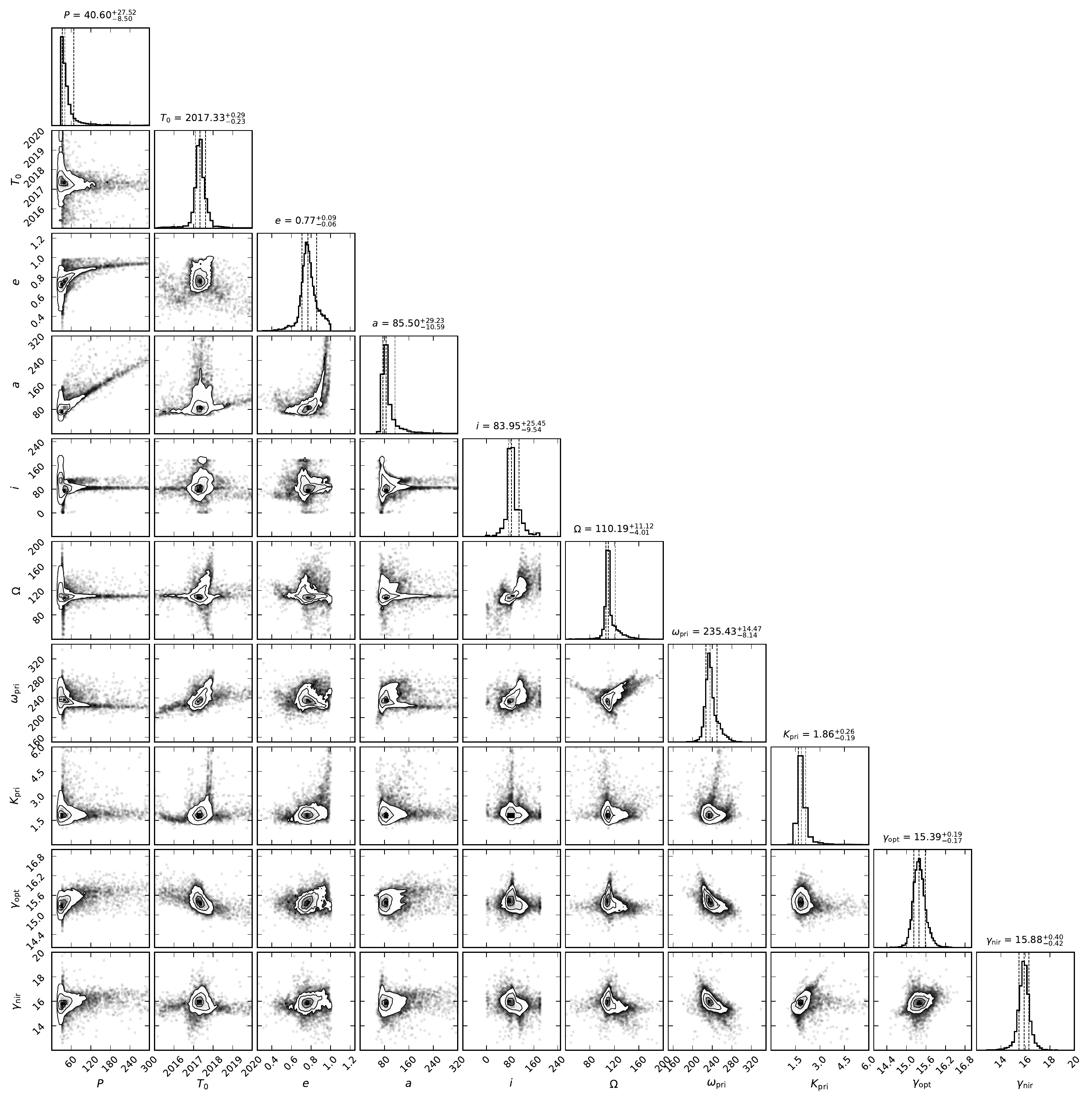}
\caption{
    Corner plot for 10\,000 iterations of the MC bootstrap uncertainty estimate results in Section~\ref{sec:orbit_fit}.
    The units for each parameter are: 
    $P$ in days, $T_0$ in Julian year, $a$ in AU, $i$ in deg, $\Omega$ in deg,
    $\omega_{\rm pri}$ in deg, $K_{\rm pri}$ in \kms, $\gamma_{\rm opt}$ in \kms,
    and $\gamma_{\rm nir}$ in \kms.
    }
\label{fig:corner_orb}
\end{figure*}

Figure~\ref{fig:corner_orb} shows the corner plot of the 10\,000 iterations for the MC bootstrap results described in Section~\ref{sec:orbit_fit}.

\clearpage
\bibliography{main}

\begin{thebibliography}{}
\expandafter\ifx\csname natexlab\endcsname\relax\def\natexlab#1{#1}\fi
\providecommand{\url}[1]{\href{#1}{#1}}
\providecommand{\dodoi}[1]{doi:~\href{http://doi.org/#1}{\nolinkurl{#1}}}
\providecommand{\doeprint}[1]{\href{http://ascl.net/#1}{\nolinkurl{http://ascl.net/#1}}}
\providecommand{\doarXiv}[1]{\href{https://arxiv.org/abs/#1}{\nolinkurl{https://arxiv.org/abs/#1}}}

\bibitem[{Allard \& Hauschildt(1995)}]{allard1995}
Allard, F., \& Hauschildt, P.~H. 1995, M (Sub) {{Dwarf Model Atmospheres}}:
  {{The Next Generation}}, 32.
\newblock \url{https://ui.adsabs.harvard.edu/abs/1995bmsb.conf...32A}

\bibitem[{Allard {et~al.}(1997)Allard, Hauschildt, Alexander, \&
  Starrfield}]{allard1997}
Allard, F., Hauschildt, P.~H., Alexander, D.~R., \& Starrfield, S. 1997, Annual
  Review of Astronomy and Astrophysics, 35, 137,
  \dodoi{10.1146/annurev.astro.35.1.137}

\bibitem[{Allard {et~al.}(2011)Allard, Homeier, \& Freytag}]{allard2011}
Allard, F., Homeier, D., \& Freytag, B. 2011, 448, 91.
\newblock \url{https://ui.adsabs.harvard.edu/abs/2011ASPC..448...91A}

\bibitem[{Allen {et~al.}(2017)Allen, Prato, {Wright-Garba}, Schaefer, Biddle,
  Skiff, Avilez, Muzzio, \& Simon}]{allen2017}
Allen, T.~S., Prato, L., {Wright-Garba}, N., {et~al.} 2017, The Astrophysical
  Journal, 845, 161, \dodoi{10.3847/1538-4357/aa8094}

\bibitem[{{Astropy Collaboration} {et~al.}(2013){Astropy Collaboration},
  Robitaille, Tollerud, Greenfield, Droettboom, Bray, Aldcroft, Davis,
  Ginsburg, {Price-Whelan}, Kerzendorf, Conley, Crighton, Barbary, Muna,
  Ferguson, Grollier, Parikh, Nair, Unther, Deil, Woillez, Conseil, Kramer,
  Turner, Singer, Fox, Weaver, Zabalza, Edwards, Azalee~Bostroem, Burke, Casey,
  Crawford, Dencheva, Ely, Jenness, Labrie, Lim, Pierfederici, Pontzen, Ptak,
  Refsdal, Servillat, \& Streicher}]{astropycollaboration2013}
{Astropy Collaboration}, Robitaille, T.~P., Tollerud, E.~J., {et~al.} 2013,
  Astronomy and Astrophysics, 558, A33, \dodoi{10.1051/0004-6361/201322068}

\bibitem[{{Astropy Collaboration} {et~al.}(2018){Astropy Collaboration},
  {Price-Whelan}, Sip{\H o}cz, G{\"u}nther, Lim, Crawford, Conseil, Shupe,
  Craig, Dencheva, Ginsburg, VanderPlas, Bradley, {P{\'e}rez-Su{\'a}rez}, {de
  Val-Borro}, Aldcroft, Cruz, Robitaille, Tollerud, Ardelean, Babej, Bach,
  Bachetti, Bakanov, Bamford, Barentsen, Barmby, Baumbach, Berry, Biscani,
  Boquien, Bostroem, Bouma, Brammer, Bray, Breytenbach, Buddelmeijer, Burke,
  Calderone, Cano~Rodr{\'i}guez, Cara, Cardoso, Cheedella, Copin, Corrales,
  Crichton, D'Avella, Deil, Depagne, Dietrich, Donath, Droettboom, Earl, Erben,
  Fabbro, Ferreira, Finethy, Fox, Garrison, Gibbons, Goldstein, Gommers, Greco,
  Greenfield, Groener, Grollier, Hagen, Hirst, Homeier, Horton, Hosseinzadeh,
  Hu, Hunkeler, Ivezi{\'c}, Jain, Jenness, Kanarek, Kendrew, Kern, Kerzendorf,
  Khvalko, King, Kirkby, Kulkarni, Kumar, Lee, Lenz, Littlefair, Ma, Macleod,
  Mastropietro, McCully, Montagnac, Morris, Mueller, Mumford, Muna, Murphy,
  Nelson, Nguyen, Ninan, N{\"o}the, Ogaz, Oh, Parejko, Parley, Pascual, Patil,
  Patil, Plunkett, Prochaska, Rastogi, Reddy~Janga, Sabater, Sakurikar,
  Seifert, Sherbert, {Sherwood-Taylor}, Shih, Sick, Silbiger, Singanamalla,
  Singer, Sladen, Sooley, Sornarajah, Streicher, Teuben, Thomas, Tremblay,
  Turner, Terr{\'o}n, {van Kerkwijk}, {de la Vega}, Watkins, Weaver, Whitmore,
  Woillez, Zabalza, \& {Astropy Contributors}}]{astropycollaboration2018}
{Astropy Collaboration}, {Price-Whelan}, A.~M., Sip{\H o}cz, B.~M., {et~al.}
  2018, The Astronomical Journal, 156, 123, \dodoi{10.3847/1538-3881/aabc4f}

\bibitem[{{Bailer-Jones} {et~al.}(2021){Bailer-Jones}, Rybizki, Fouesneau,
  Demleitner, \& Andrae}]{bailer-jones2021}
{Bailer-Jones}, C. A.~L., Rybizki, J., Fouesneau, M., Demleitner, M., \&
  Andrae, R. 2021, The Astronomical Journal, 161, 147,
  \dodoi{10.3847/1538-3881/abd806}

\bibitem[{Baraffe {et~al.}(2015)Baraffe, Homeier, Allard, \&
  Chabrier}]{baraffe2015}
Baraffe, I., Homeier, D., Allard, F., \& Chabrier, G. 2015, A\&A, 577, A42,
  \dodoi{10.1051/0004-6361/201425481}

\bibitem[{Bellm {et~al.}(2019)Bellm, Kulkarni, Graham, Dekany, Smith, Riddle,
  Masci, Helou, Prince, Adams, Barbarino, Barlow, Bauer, Beck, Belicki, Biswas,
  Blagorodnova, Bodewits, Bolin, Brinnel, Brooke, Bue, Bulla, Burruss, Cenko,
  Chang, Connolly, Coughlin, Cromer, Cunningham, De, Delacroix, Desai, Duev,
  Eadie, Farnham, Feeney, Feindt, Flynn, Franckowiak, Frederick, Fremling,
  {Gal-Yam}, Gezari, Giomi, Goldstein, Golkhou, Goobar, Groom, Hacopians, Hale,
  Henning, Ho, Hover, Howell, Hung, Huppenkothen, Imel, Ip, Ivezi{\'c},
  Jackson, Jones, Juric, Kasliwal, Kaspi, Kaye, Kelley, Kowalski, Kramer,
  Kupfer, Landry, Laher, Lee, Lin, Lin, Lunnan, Giomi, Mahabal, Mao, Miller,
  Monkewitz, Murphy, Ngeow, Nordin, Nugent, Ofek, Patterson, Penprase, Porter,
  Rauch, Rebbapragada, Reiley, Rigault, Rodriguez, {van Roestel}, Rusholme,
  {van Santen}, Schulze, Shupe, Singer, Soumagnac, Stein, Surace, Sollerman,
  Szkody, Taddia, Terek, Van~Sistine, {van Velzen}, Vestrand, Walters, Ward,
  Ye, Yu, Yan, \& Zolkower}]{bellm2019}
Bellm, E.~C., Kulkarni, S.~R., Graham, M.~J., {et~al.} 2019, Publications of
  the Astronomical Society of the Pacific, 131, 018002,
  \dodoi{10.1088/1538-3873/aaecbe}

\bibitem[{Bertout {et~al.}(1988)Bertout, Basri, \& Bouvier}]{bertout1988}
Bertout, C., Basri, G., \& Bouvier, J. 1988, ApJ, 330, 350,
  \dodoi{10.1086/166476}

\bibitem[{Biller {et~al.}(2022)Biller, Grandjean, Messina, Desidera, Delorme,
  Lagrange, Hambsch, Mesa, Janson, Gratton, D'Orazi, Langlois, Maire,
  Schlieder, Henning, Zurlo, Hagelberg, {Brown-Sevilla}, Romero, Bonnefoy,
  Chauvin, Feldt, Meyer, Vigan, Pavlov, Soenke, LeMignant, \&
  Roux}]{biller2022}
Biller, B.~A., Grandjean, A., Messina, S., {et~al.} 2022, Astronomy and
  Astrophysics, 658, A145, \dodoi{10.1051/0004-6361/202142438}

\bibitem[{Bouvier {et~al.}(2007)Bouvier, Alencar, Harries, {Johns-Krull}, \&
  Romanova}]{bouvier2007}
Bouvier, J., Alencar, S. H.~P., Harries, T.~J., {Johns-Krull}, C.~M., \&
  Romanova, M.~M. 2007, Magnetospheric {{Accretion}} in {{Classical T Tauri
  Stars}} ({eprint: arXiv:astro-ph/0603498}), 479.
\newblock \url{https://ui.adsabs.harvard.edu/abs/2007prpl.conf..479B}

\bibitem[{Bouvier {et~al.}(1993)Bouvier, Cabrit, Fernandez, Martin, \&
  Matthews}]{bouvier1993}
Bouvier, J., Cabrit, S., Fernandez, M., Martin, E.~L., \& Matthews, J.~M. 1993,
  Astronomy and Astrophysics, 272, 176.
\newblock \url{https://ui.adsabs.harvard.edu/abs/1993A&A...272..176B}

\bibitem[{Box(1965)}]{box1965}
Box, M.~J. 1965, The Computer Journal, 8, 42, \dodoi{10.1093/comjnl/8.1.42}

\bibitem[{Cardelli {et~al.}(1989)Cardelli, Clayton, \& Mathis}]{cardelli1989}
Cardelli, J.~A., Clayton, G.~C., \& Mathis, J.~S. 1989, 135, 5.
\newblock \url{https://ui.adsabs.harvard.edu/abs/1989IAUS..135P...5C}

\bibitem[{Chen {et~al.}(1990)Chen, Simon, Longmore, Howell, \&
  Benson}]{chen1990}
Chen, W.~P., Simon, M., Longmore, A.~J., Howell, R.~R., \& Benson, J.~A. 1990,
  ApJ, 357, 224, \dodoi{10.1086/168908}

\bibitem[{Conroy {et~al.}(2020)Conroy, Kochoska, Hey, Pablo, Hambleton, Jones,
  Giammarco, {Abdul-Masih}, \& Pr{\v s}a}]{conroy2020}
Conroy, K.~E., Kochoska, A., Hey, D., {et~al.} 2020, ApJS, 250, 34,
  \dodoi{10.3847/1538-4365/abb4e2}

\bibitem[{Crockett {et~al.}(2012)Crockett, Mahmud, Prato, {Johns-Krull}, Jaffe,
  Hartigan, \& Beichman}]{crockett2012}
Crockett, C.~J., Mahmud, N.~I., Prato, L., {et~al.} 2012, ApJ, 761, 164,
  \dodoi{10.1088/0004-637X/761/2/164}

\bibitem[{Cutri \& {et al.}(2012)}]{cutri2012}
Cutri, R.~M., \& {et al.} 2012, VizieR Online Data Catalog, II/311.
\newblock \url{https://ui.adsabs.harvard.edu/abs/2012yCat.2311....0C}

\bibitem[{Cutri {et~al.}(2021)Cutri, Wright, Conrow, Fowler, Eisenhardt,
  Grillmair, Kirkpatrick, Masci, McCallon, Wheelock, {Fajardo-Acosta}, Yan,
  Benford, Harbut, Jarrett, Lake, Leisawitz, Ressler, Stanford, Tsai, Liu,
  Helou, Mainzer, Gettngs, Gonzalez, Hoffman, Marsh, Padgett, Skrutskie, Beck,
  Papin, \& Wittman}]{cutri2021}
Cutri, R.~M., Wright, E.~L., Conrow, T., {et~al.} 2021, VizieR Online Data
  Catalog, II/328.
\newblock \url{https://ui.adsabs.harvard.edu/abs/2014yCat.2328....0C}

\bibitem[{Czekala {et~al.}(2021)Czekala, Ribas, Cuello, Chiang, Mac{\'i}as,
  Duch{\^e}ne, Andrews, \& Espaillat}]{czekala2021}
Czekala, I., Ribas, {\'A}., Cuello, N., {et~al.} 2021, ApJ, 912, 6,
  \dodoi{10.3847/1538-4357/abebe3}

\bibitem[{Daemgen {et~al.}(2015)Daemgen, Bonavita, Jayawardhana,
  Lafreni{\`e}re, \& Janson}]{daemgen2015}
Daemgen, S., Bonavita, M., Jayawardhana, R., Lafreni{\`e}re, D., \& Janson, M.
  2015, ApJ, 799, 155, \dodoi{10.1088/0004-637X/799/2/155}

\bibitem[{Deen(2013)}]{deen2013}
Deen, C.~P. 2013, The Astronomical Journal, 146, 51,
  \dodoi{10.1088/0004-6256/146/3/51}

\bibitem[{Evans {et~al.}(2003)Evans, Allen, Blake, Boogert, Bourke, Harvey,
  Kessler, Koerner, Lee, Mundy, Myers, Padgett, Pontoppidan, Sargent,
  Stapelfeldt, {van Dishoeck}, Young, \& Young}]{evans2003}
Evans, II, N.~J., Allen, L.~E., Blake, G.~A., {et~al.} 2003, Publications of
  the Astronomical Society of the Pacific, 115, 965, \dodoi{10.1086/376697}

\bibitem[{Eyer {et~al.}(2022)Eyer, Audard, Holl, Rimoldini, Carnerero,
  Clementini, De~Ridder, Distefano, Evans, Gavras, Gomel, Lebzelter, Marton,
  Mowlavi, Panahi, Ripepi, Wyrzykowski, Nienartowicz, {Jevardat de Fombelle},
  {Lecoeur-Taibi}, Rohrbasser, Riello, {Garcia-Lario}, Lanzafame, Mazeh,
  Raiteri, Zucker, Abraham, Aerts, Aguado, Anderson, Bashi, Binnenfeld,
  Faigler, Garofalo, Karbevska, Kospal, Kruszynska, Kun, Lanza, Leccia,
  Marconi, Messina, Molinaro, Molnar, Muraveva, Musella, Nagy, Pagano,
  Palaversa, Plachy, Rybicki, Shahaf, Szabados, {Szegedi-Elek}, Trabucchi,
  Barblan, \& Roelens}]{eyer2022}
Eyer, L., Audard, M., Holl, B., {et~al.} 2022, Gaia {{Data Release}} 3.
  {{Summary}} of the Variability Processing and Analysis.
\newblock \url{https://ui.adsabs.harvard.edu/abs/2022arXiv220606416E}

\bibitem[{Feiden(2016)}]{feiden2016}
Feiden, G.~A. 2016, A\&A, 593, A99, \dodoi{10.1051/0004-6361/201527613}

\bibitem[{Flores {et~al.}(2019)Flores, Connelley, Reipurth, \&
  Boogert}]{flores2019}
Flores, C., Connelley, M.~S., Reipurth, B., \& Boogert, A. 2019, The
  Astrophysical Journal, 882, 75, \dodoi{10.3847/1538-4357/ab35d4}

\bibitem[{Flores {et~al.}(2022)Flores, Connelley, Reipurth, \&
  Duch{\^e}ne}]{flores2022}
Flores, C., Connelley, M.~S., Reipurth, B., \& Duch{\^e}ne, G. 2022, ApJ, 925,
  21, \dodoi{10.3847/1538-4357/ac37bd}

\bibitem[{Folha \& Emerson(1999)}]{folha1999}
Folha, D. F.~M., \& Emerson, J.~P. 1999, A\&A, 15

\bibitem[{{Gaia Collaboration} {et~al.}(2022){Gaia Collaboration}, Vallenari,
  Brown, Prusti, {de Bruijne}, Arenou, Babusiaux, Biermann, Creevey, Ducourant,
  Evans, Eyer, Guerra, Hutton, Jordi, Klioner, Lammers, Lindegren, Luri,
  Mignard, Panem, Pourbaix, Randich, Sartoretti, Soubiran, Tanga, Walton,
  {Bailer-Jones}, Bastian, Drimmel, Jansen, Katz, Lattanzi, {van Leeuwen},
  Bakker, Cacciari, Casta{\~n}eda, De~Angeli, Fabricius, Fouesneau, Fr{\'e}mat,
  Galluccio, Guerrier, Heiter, Masana, Messineo, Mowlavi, Nicolas,
  Nienartowicz, Pailler, Panuzzo, Riclet, Roux, Seabroke, Sordo{\o}rcit,
  Th{\'e}venin, {Gracia-Abril}, Portell, Teyssier, Altmann, Andrae, Audard,
  {Bellas-Velidis}, Benson, Berthier, Blomme, Burgess, Busonero, Busso,
  C{\'a}novas, Carry, Cellino, Cheek, Clementini, Damerdji, Davidson, {de
  Teodoro}, Nu{\~n}ez~Campos, Delchambre, Dell'Oro, Esquej,
  {Fern{\'a}ndez-Hern{\'a}ndez}, Fraile, Garabato, {Garc{\'i}a-Lario}, Gosset,
  Haigron, Halbwachs, Hambly, Harrison, Hern{\'a}ndez, Hestroffer, Hodgkin,
  Holl, Jan{\ss}en, {Jevardat de Fombelle}, Jordan, {Krone-Martins}, Lanzafame,
  L{\"o}ffler, Marchal, Marrese, Moitinho, Muinonen, Osborne, Pancino, Pauwels,
  {Recio-Blanco}, Reyl{\'e}, Riello, Rimoldini, Roegiers, Rybizki, Sarro,
  Siopis, Smith, Sozzetti, Utrilla, {van Leeuwen}, Abbas, {\'A}brah{\'a}m,
  Abreu~Aramburu, Aerts, Aguado, Ajaj, {Aldea-Montero}, Altavilla, {\'A}lvarez,
  Alves, Anders, Anderson, Anglada~Varela, Antoja, Baines, Baker,
  {Balaguer-N{\'u}{\~n}ez}, Balbinot, Balog, Barache, Barbato, Barros, Barstow,
  Bartolom{\'e}, Bassilana, Bauchet, Becciani, Bellazzini, Berihuete, Bernet,
  Bertone, Bianchi, Binnenfeld, {Blanco-Cuaresma}, Blazere, Boch, Bombrun,
  Bossini, Bouquillon, Bragaglia, Bramante, Breedt, Bressan, Brouillet,
  Brugaletta, Bucciarelli, Burlacu, Butkevich, Buzzi, Caffau, Cancelliere,
  {Cantat-Gaudin}, Carballo, Carlucci, Carnerero, Carrasco, Casamiquela,
  Castellani, {Castro-Ginard}, Chaoul, Charlot, Chemin, Chiaramida, Chiavassa,
  Chornay, Comoretto, Contursi, Cooper, Cornez, Cowell, Crifo, Cropper, Crosta,
  Crowley, Dafonte, Dapergolas, David, David, {de Laverny}, De~Luise, De~March,
  De~Ridder, {de Souza}, {de Torres}, {del Peloso}, {del Pozo}, Delbo, Delgado,
  Delisle, Demouchy, Dharmawardena, Di~Matteo, Diakite, Diener, Distefano,
  Dolding, Edvardsson, Enke, Fabre, Fabrizio, Faigler, Fedorets, Fernique,
  Fienga, Figueras, Fournier, Fouron, Fragkoudi, Gai, {Garcia-Gutierrez},
  {Garcia-Reinaldos}, {Garc{\'i}a-Torres}, Garofalo, Gavel, Gavras, Gerlach,
  Geyer, Giacobbe, Gilmore, Girona, Giuffrida, Gomel, Gomez,
  {Gonz{\'a}lez-N{\'u}{\~n}ez}, {Gonz{\'a}lez-Santamar{\'i}a},
  {Gonz{\'a}lez-Vidal}, Granvik, Guillout, Guiraud,
  {Guti{\'e}rrez-S{\'a}nchez}, Guy, Hatzidimitriou, Hauser, Haywood, Helmer,
  Helmi, Sarmiento, Hidalgo, Hilger, H{\l}adczuk, Hobbs, Holland, Huckle,
  Jardine, Jasniewicz, {Jean-Antoine Piccolo}, {Jim{\'e}nez-Arranz}, Jorissen,
  Juaristi~Campillo, Julbe, Karbevska, Kervella, Khanna, Kontizas, Kordopatis,
  Korn, K{\'o}sp{\'a}l, {Kostrzewa-Rutkowska}, Kruszy{\'n}ska, Kun, Laizeau,
  Lambert, Lanza, Lasne, Le~Campion, Lebreton, Lebzelter, Leccia, Leclerc,
  {Lecoeur-Taibi}, Liao, Licata, Lindstr{\o}m, Lister, Livanou, Lobel, Lorca,
  Loup, Madrero~Pardo, Magdaleno~Romeo, Managau, Mann, Manteiga, Marchant,
  Marconi, Marcos, Marcos~Santos, Mar{\'i}n~Pina, Marinoni, Marocco, Marshall,
  Polo, {Mart{\'i}n-Fleitas}, Marton, Mary, Masip, Massari,
  {Mastrobuono-Battisti}, Mazeh, McMillan, Messina, Michalik, Millar, Mints,
  Molina, Molinaro, Moln{\'a}r, Monari, Mongui{\'o}, Montegriffo, Montero, Mor,
  Mora, Morbidelli, Morel, Morris, Muraveva, Murphy, Musella, Nagy, Noval,
  Oca{\~n}a, Ogden, Ordenovic, Osinde, Pagani, Pagano, Palaversa, Palicio,
  {Pallas-Quintela}, Panahi, {Payne-Wardenaar}, Pe{\~n}alosa~Esteller,
  Penttil{\"a}, Pichon, Piersimoni, Pineau, Plachy, Plum, Poggio, Pr{\v s}a,
  Pulone, Racero, Ragaini, Rainer, Raiteri, Rambaux, Ramos, {Ramos-Lerate},
  Re~Fiorentin, Regibo, Richards, Rios~Diaz, Ripepi, Riva, Rix, Rixon,
  Robichon, Robin, Robin, Roelens, Rogues, Rohrbasser, {Romero-G{\'o}mez},
  Rowell, Royer, Ruz~Mieres, Rybicki, Sadowski, S{\'a}ez~N{\'u}{\~n}ez,
  Sagrist{\`a}~Sell{\'e}s, Sahlmann, Salguero, Samaras, Sanchez~Gimenez, Sanna,
  Santove{\~n}a, Sarasso, Schultheis, Sciacca, Segol, Segovia, S{\'e}gransan,
  Semeux, Shahaf, Siddiqui, Siebert, Siltala, Silvelo, Slezak, Slezak, Smart,
  Snaith, Solano, Solitro, Souami, Souchay, Spagna, Spina, Spoto, Steele,
  Steidelm{\"u}ller, Stephenson, S{\"u}veges, Surdej, Szabados, {Szegedi-Elek},
  Taris, Taylo, Teixeira, Tolomei, Tonello, Torra, Torra, Torralba~Elipe,
  Trabucchi, Tsounis, Turon, Ulla, Unger, Vaillant, {van Dillen}, {van Reeven},
  Vanel, Vecchiato, Viala, Vicente, Voutsinas, Weiler, Wevers, Wyrzykowski,
  Yoldas, Yvard, Zhao, Zorec, Zucker, \& Zwitter}]{gaiacollaboration2022}
{Gaia Collaboration}, Vallenari, A., Brown, A. G.~A., {et~al.} 2022, Gaia
  {{Data Release}} 3: {{Summary}} of the Content and Survey Properties.
\newblock \url{https://ui.adsabs.harvard.edu/abs/2022arXiv220800211G}

\bibitem[{Ghez {et~al.}(1993)Ghez, Neugebauer, \& Matthews}]{ghez1993a}
Ghez, A.~M., Neugebauer, G., \& Matthews, K. 1993, AJ, 106, 2005,
  \dodoi{10.1086/116782}

\bibitem[{Grankin {et~al.}(2008)Grankin, Bouvier, Herbst, \&
  Melnikov}]{grankin2008}
Grankin, K.~N., Bouvier, J., Herbst, W., \& Melnikov, {\relax S. Yu}. 2008,
  Astronomy and Astrophysics, 479, 827, \dodoi{10.1051/0004-6361:20078476}

\bibitem[{Grankin {et~al.}(2007)Grankin, Melnikov, Bouvier, Herbst, \&
  Shevchenko}]{grankin2007}
Grankin, K.~N., Melnikov, {\relax S. Yu}., Bouvier, J., Herbst, W., \&
  Shevchenko, V.~S. 2007, Astronomy and Astrophysics, 461, 183,
  \dodoi{10.1051/0004-6361:20065489}

\bibitem[{Gray(2005)}]{gray2005}
Gray, D.~F. 2005, The {{Observation}} and {{Analysis}} of {{Stellar
  Photospheres}}.
\newblock \url{https://ui.adsabs.harvard.edu/abs/2005oasp.book.....G}

\bibitem[{Gullikson {et~al.}(2014)Gullikson, {Dodson-Robinson}, \&
  Kraus}]{gullikson2014}
Gullikson, K., {Dodson-Robinson}, S., \& Kraus, A. 2014, AJ, 148, 53,
  \dodoi{10.1088/0004-6256/148/3/53}

\bibitem[{Gustafsson {et~al.}(2008)Gustafsson, Edvardsson, Eriksson,
  J{\o}rgensen, Nordlund, \& Plez}]{gustafsson2008}
Gustafsson, B., Edvardsson, B., Eriksson, K., {et~al.} 2008, Astronomy and
  Astrophysics, 486, 951, \dodoi{10.1051/0004-6361:200809724}

\bibitem[{Harris {et~al.}(2020)Harris, Millman, {van der Walt}, Gommers,
  Virtanen, Cournapeau, Wieser, Taylor, Berg, Smith, Kern, Picus, Hoyer, {van
  Kerkwijk}, Brett, Haldane, {del R{\'i}o}, Wiebe, Peterson,
  {G{\'e}rard-Marchant}, Sheppard, Reddy, Weckesser, Abbasi, Gohlke, \&
  Oliphant}]{harris2020}
Harris, C.~R., Millman, K.~J., {van der Walt}, S.~J., {et~al.} 2020, Nature,
  585, 357, \dodoi{10.1038/s41586-020-2649-2}

\bibitem[{Hartmann {et~al.}(2016)Hartmann, Herczeg, \& Calvet}]{hartmann2016}
Hartmann, L., Herczeg, G., \& Calvet, N. 2016, Annu. Rev. Astron. Astrophys.,
  54, 135, \dodoi{10.1146/annurev-astro-081915-023347}

\bibitem[{Hartmann {et~al.}(2005)Hartmann, Megeath, Allen, Luhman, Calvet,
  D'Alessio, Franco-Hernandez, \& Fazio}]{hartmann2005}
Hartmann, L., Megeath, S.~T., Allen, L., {et~al.} 2005, ApJ, 629, 881,
  \dodoi{10.1086/431472}

\bibitem[{Hauschildt {et~al.}(1999)Hauschildt, Allard, Ferguson, Baron, \&
  Alexander}]{hauschildt1999}
Hauschildt, P.~H., Allard, F., Ferguson, J., Baron, E., \& Alexander, D.~R.
  1999, The Astrophysical Journal, 525, 871, \dodoi{10.1086/307954}

\bibitem[{Hawley \& Fisher(1994)}]{hawley1994}
Hawley, S.~L., \& Fisher, G.~H. 1994, The Astrophysical Journal, 426, 387,
  \dodoi{10.1086/174075}

\bibitem[{Herbst {et~al.}(1994)Herbst, Herbst, Grossman, \&
  Weinstein}]{herbst1994}
Herbst, W., Herbst, D.~K., Grossman, E.~J., \& Weinstein, D. 1994, The
  Astronomical Journal, 108, 1906, \dodoi{10.1086/117204}

\bibitem[{Herczeg \& Hillenbrand(2014)}]{herczeg2014}
Herczeg, G.~J., \& Hillenbrand, L.~A. 2014, ApJ, 786, 97,
  \dodoi{10.1088/0004-637X/786/2/97}

\bibitem[{Hillenbrand \& White(2004)}]{hillenbrand2004}
Hillenbrand, L.~A., \& White, R.~J. 2004, ApJ, 604, 741, \dodoi{10.1086/382021}

\bibitem[{Hinkle {et~al.}(2000)Hinkle, Joyce, Sharp, \& Valenti}]{hinkle2000}
Hinkle, K.~H., Joyce, R.~R., Sharp, N., \& Valenti, J.~A. 2000, 4008, 720,
  \dodoi{10.1117/12.395529}

\bibitem[{Hunter(2007)}]{hunter2007}
Hunter, J.~D. 2007, Comput. Sci. Eng., 9, 90, \dodoi{10.1109/MCSE.2007.55}

\bibitem[{Husser {et~al.}(2013)Husser, {Wende-von Berg}, Dreizler, Homeier,
  Reiners, Barman, \& Hauschildt}]{husser2013}
Husser, T.-O., {Wende-von Berg}, S., Dreizler, S., {et~al.} 2013, A\&A, 553,
  A6, \dodoi{10.1051/0004-6361/201219058}

\bibitem[{Itoh {et~al.}(2008)Itoh, Tamura, Hayashi, Oasa, Hayashi, Fukagawa,
  Kudo, Mayama, Ishii, Pyo, Yamashita, \& Morino}]{itoh2008}
Itoh, Y., Tamura, M., Hayashi, M., {et~al.} 2008, PASJ, 60, 209,
  \dodoi{10.1093/pasj/60.2.209}

\bibitem[{Johns-Krull(2007)}]{johns-krull2007}
Johns-Krull, C.~M. 2007, ApJ, 664, 975, \dodoi{10.1086/519017}

\bibitem[{Johns-Krull {et~al.}(2004)Johns-Krull, Valenti, \&
  Saar}]{johns-krull2004}
Johns-Krull, C.~M., Valenti, J.~A., \& Saar, S.~H. 2004, ApJ, 617, 1204,
  \dodoi{10.1086/425652}

\bibitem[{{Johns-Krull} {et~al.}(2016){Johns-Krull}, McLane, Prato, Crockett,
  Jaffe, Hartigan, Beichman, Mahmud, Chen, Skiff, Cauley, Jones, \&
  Mace}]{johns-krull2016}
{Johns-Krull}, C.~M., McLane, J.~N., Prato, L., {et~al.} 2016, ApJ, 826, 206,
  \dodoi{10.3847/0004-637X/826/2/206}

\bibitem[{Johnson \& Morgan(1953)}]{johnson1953}
Johnson, H.~L., \& Morgan, W.~W. 1953, The Astrophysical Journal, 117, 313,
  \dodoi{10.1086/145697}

\bibitem[{Johnson {et~al.}(2008)Johnson, Marcy, Fischer, Wright, Reffert,
  Kregenow, Williams, \& Peek}]{johnson2008}
Johnson, J.~A., Marcy, G.~W., Fischer, D.~A., {et~al.} 2008, ApJ, 675, 784,
  \dodoi{10.1086/526453}

\bibitem[{Katz {et~al.}(2022)Katz, Sartoretti, Guerrier, Panuzzo, Seabroke,
  Th{\'e}venin, Cropper, Benson, Blomme, Haigron, Marchal, Smith, Baker,
  Chemin, Damerdji, David, Dolding, Fr{\'e}mat, Gosset, Jan{\ss}en, Jasniewicz,
  Lobel, Plum, Samaras, Snaith, Soubiran, Vanel, Zwitter, Antoja, Arenou,
  Babusiaux, Brouillet, Caffau, Di~Matteo, Fabre, Fabricius, Frakgoudi,
  Haywood, Huckle, Hottier, Lasne, Leclerc, {Mastrobuono-Battisti}, Royer,
  Teyssier, Zorec, Crifo, {Jean-Antoine Piccolo}, Turon, \& Viala}]{katz2022}
Katz, D., Sartoretti, P., Guerrier, A., {et~al.} 2022, Gaia {{Data Release}} 3
  {{Properties}} and Validation of the Radial Velocities.
\newblock \url{https://ui.adsabs.harvard.edu/abs/2022arXiv220605902K}

\bibitem[{Kenyon \& Hartmann(1995)}]{kenyon1995}
Kenyon, S.~J., \& Hartmann, L. 1995, ApJS, 101, 117, \dodoi{10.1086/192235}

\bibitem[{Kochukhov(2007)}]{kochukhov2007}
Kochukhov, O. 2007, arXiv:astro-ph/0701084.
\newblock \doeprint{astro-ph/0701084}

\bibitem[{Kraus {et~al.}(2011)Kraus, Ireland, Martinache, \&
  Hillenbrand}]{kraus2011}
Kraus, A.~L., Ireland, M.~J., Martinache, F., \& Hillenbrand, L.~A. 2011, ApJ,
  731, 8, \dodoi{10.1088/0004-637X/731/1/8}

\bibitem[{Kupka {et~al.}(1999)Kupka, Piskunov, Ryabchikova, Stempels, \&
  Weiss}]{kupka1999}
Kupka, F., Piskunov, N., Ryabchikova, T.~A., Stempels, H.~C., \& Weiss, W.~W.
  1999, Astronomy and Astrophysics Supplement Series, 138, 119,
  \dodoi{10.1051/aas:1999267}

\bibitem[{Lee {et~al.}(2017)Lee, Gullikson, \&
  Kaplan}]{jae_joon_lee_2017_845059}
Lee, J.-J., Gullikson, K., \& Kaplan, K. 2017, Igrins/Plp 2.2.0, Zenodo,
  \dodoi{10.5281/zenodo.845059}

\bibitem[{Levine {et~al.}(2018)Levine, Mace, Jaffe, Sokal, Lee, Oh, Park,
  Kaplan, Yuk, Chun, Jeong, Pak, Kim, Lee, Good, Kidder, Oh, Lee, Yu, Hwang,
  Park, Kim, Chinn, Peck, Diaz, Rutten, Prato, Jacoby, Nofi, Hardesty, DeGroff,
  Cornelius, Dunham, Nah, {Lopez-Valdivia}, MacQueen, \&
  Weinberger}]{levine2018a}
Levine, S., Mace, G.~N., Jaffe, D.~T., {et~al.} 2018, in Ground-Based and
  {{Airborne Instrumentation}} for {{Astronomy VII}}, ed. H.~Takami, C.~J.
  Evans, \& L.~Simard ({Austin, United States}: {SPIE}), 26,
  \dodoi{10.1117/12.2312345}

\bibitem[{Lindegren {et~al.}(2021)Lindegren, Klioner, Hern{\'a}ndez, Bombrun,
  {Ramos-Lerate}, Steidelm{\"u}ller, Bastian, Biermann, {de Torres}, Gerlach,
  Geyer, Hilger, Hobbs, Lammers, McMillan, Stephenson, Casta{\~n}eda, Davidson,
  Fabricius, {Gracia-Abril}, Portell, Rowell, Teyssier, Torra, Bartolom{\'e},
  Clotet, Garralda, {Gonz{\'a}lez-Vidal}, Torra, Abbas, Altmann,
  Anglada~Varela, {Balaguer-N{\'u}{\~n}ez}, Balog, Barache, Becciani, Bernet,
  Bertone, Bianchi, Bouquillon, Brown, Bucciarelli, Busonero, Butkevich, Buzzi,
  Cancelliere, Carlucci, Charlot, Cioni, Crosta, Crowley, {del Peloso}, {del
  Pozo}, Drimmel, Esquej, Fienga, Fraile, Gai, {Garcia-Reinaldos}, Guerra,
  Hambly, Hauser, Jan{\ss}en, Jordan, {Kostrzewa-Rutkowska}, Lattanzi, Liao,
  Licata, Lister, L{\"o}ffler, Marchant, Masip, Mignard, Mints, Molina, Mora,
  Morbidelli, Murphy, Pagani, Panuzzo, Pe{\~n}alosa~Esteller, Poggio,
  Re~Fiorentin, Riva, Sagrist{\`a}~Sell{\'e}s, Sanchez~Gimenez, Sarasso,
  Sciacca, Siddiqui, Smart, Souami, Spagna, Steele, Taris, Utrilla, {van
  Reeven}, \& Vecchiato}]{lindegren2021a}
Lindegren, L., Klioner, S.~A., Hern{\'a}ndez, J., {et~al.} 2021, Astronomy and
  Astrophysics, 649, A2, \dodoi{10.1051/0004-6361/202039709}

\bibitem[{Livingston \& Wallace(1991)}]{livingston1991}
Livingston, W., \& Wallace, L. 1991, An Atlas of the Solar Spectrum in the
  Infrared from 1850 to 9000 Cm-1 (1.1 to 5.4 Micrometer).
\newblock \url{https://ui.adsabs.harvard.edu/abs/1991aass.book.....L}

\bibitem[{{L{\'o}pez-Valdivia} {et~al.}(2021){L{\'o}pez-Valdivia}, Sokal, Mace,
  Kidder, Hussaini, Nofi, Prato, {Johns-Krull}, Oh, Lee, Park, Oh, Kraus,
  Kaplan, Llama, Mann, Kim, {Gully-Santiago}, Lee, Pak, Hwang, \&
  Jaffe}]{lopez-valdivia2021}
{L{\'o}pez-Valdivia}, R., Sokal, K.~R., Mace, G.~N., {et~al.} 2021, ApJ, 921,
  53, \dodoi{10.3847/1538-4357/ac1a7b}

\bibitem[{Mace {et~al.}(2016)Mace, Jaffe, Park, \& Lee}]{mace_2016_56434}
Mace, G., Jaffe, D., Park, C., \& Lee, J.-J. 2016, Stellar Radial Velocities
  with {{IGRINS}} at {{McDonald}} Observatory,  {Zenodo},
  \dodoi{10.5281/zenodo.56434}

\bibitem[{Mahmud {et~al.}(2011)Mahmud, Crockett, {Johns-Krull}, Prato,
  Hartigan, Jaffe, \& Beichman}]{mahmud2011}
Mahmud, N.~I., Crockett, C.~J., {Johns-Krull}, C.~M., {et~al.} 2011, ApJ, 736,
  123, \dodoi{10.1088/0004-637X/736/2/123}

\bibitem[{Mann {et~al.}(2022)Mann, Wood, Schmidt, Barber, Owen, Tofflemire,
  Newton, Mamajek, Bush, Mace, Kraus, Thao, Vanderburg, Llama, {Johns-Krull},
  Prato, Stahl, Tang, Fields, Collins, Collins, Gan, Jensen, Kamler, Schwarz,
  Furlan, Gnilka, Howell, Lester, Owens, Suarez, Mekarnia, Guillot, Abe,
  Triaud, Johnson, Milburn, Rizzuto, Quinn, Kerr, Ricker, Vanderspek, Latham,
  Seager, Winn, Jenkins, Guerrero, Shporer, Schlieder, McLean, \&
  Wohler}]{mann2022}
Mann, A.~W., Wood, M.~L., Schmidt, S.~P., {et~al.} 2022, AJ, 163, 156,
  \dodoi{10.3847/1538-3881/ac511d}

\bibitem[{Mansfield {et~al.}(2022)Mansfield, Wiser, Stevenson, Smith, Line,
  Bean, Fortney, Parmentier, Kempton, Arcangeli, D{\'e}sert, Kilpatrick,
  Kreidberg, \& Malik}]{mansfield2022}
Mansfield, M., Wiser, L., Stevenson, K.~B., {et~al.} 2022, The Astronomical
  Journal, 163, 261, \dodoi{10.3847/1538-3881/ac658f}

\bibitem[{Meyer {et~al.}(1997)Meyer, Beckwith, Herbst, \& Robberto}]{meyer1997}
Meyer, M.~R., Beckwith, S. V.~W., Herbst, T.~M., \& Robberto, M. 1997, ApJ,
  489, L173, \dodoi{10.1086/310976}

\bibitem[{Nelder \& Mead(1965)}]{nelder1965}
Nelder, J.~A., \& Mead, R. 1965, The Computer Journal, 7, 308,
  \dodoi{10.1093/comjnl/7.4.308}

\bibitem[{Nguyen {et~al.}(2012)Nguyen, Brandeker, {van Kerkwijk}, \&
  Jayawardhana}]{nguyen2012}
Nguyen, D.~C., Brandeker, A., {van Kerkwijk}, M.~H., \& Jayawardhana, R. 2012,
  ApJ, 745, 119, \dodoi{10.1088/0004-637X/745/2/119}

\bibitem[{Park {et~al.}(2014)Park, Jaffe, Yuk, Chun, Pak, Kim, Pavel, Lee, Oh,
  Jeong, Sim, Lee, Nguyen~Le, Strubhar, {Gully-Santiago}, Oh, Cha, Moon, Park,
  Brooks, Ko, Han, Nah, Hill, Lee, Barnes, Yu, Kaplan, Mace, Kim, Lee, Hwang,
  \& Park}]{park2014}
Park, C., Jaffe, D.~T., Yuk, I.-S., {et~al.} 2014, in {{SPIE Astronomical
  Telescopes}} + {{Instrumentation}}, ed. S.~K. Ramsay, I.~S. McLean, \&
  H.~Takami, {Montr\'eal, Quebec, Canada}, 91471D, \dodoi{10.1117/12.2056431}

\bibitem[{Piskunov(1999)}]{piskunov1999}
Piskunov, N. 1999, 243, 515, \dodoi{10.1007/978-94-015-9329-8_45}

\bibitem[{Prato {et~al.}(2008)Prato, Huerta, {Johns-Krull}, Mahmud, Jaffe, \&
  Hartigan}]{prato2008}
Prato, L., Huerta, M., {Johns-Krull}, C.~M., {et~al.} 2008, ApJ, 687, L103,
  \dodoi{10.1086/593201}

\bibitem[{Prato {et~al.}(2002)Prato, Simon, Mazeh, McLean, Norman, \&
  Zucker}]{prato2002}
Prato, L., Simon, M., Mazeh, T., {et~al.} 2002, ApJ, 569, 863,
  \dodoi{10.1086/339397}

\bibitem[{Reiners {et~al.}(2016)Reiners, Mrotzek, Lemke, Hinrichs, \&
  Reinsch}]{reiners2016}
Reiners, A., Mrotzek, N., Lemke, U., Hinrichs, J., \& Reinsch, K. 2016, A\&A,
  587, A65, \dodoi{10.1051/0004-6361/201527530}

\bibitem[{Riello {et~al.}(2021)Riello, De~Angeli, Evans, Montegriffo, Carrasco,
  Busso, Palaversa, Burgess, Diener, Davidson, Rowell, Fabricius, Jordi,
  Bellazzini, Pancino, Harrison, Cacciari, {van Leeuwen}, Hambly, Hodgkin,
  Osborne, Altavilla, Barstow, Brown, Castellani, Cowell, De~Luise, Gilmore,
  Giuffrida, Hidalgo, Holland, Marinoni, Pagani, Piersimoni, Pulone, Ragaini,
  Rainer, Richards, Sanna, Walton, Weiler, \& Yoldas}]{riello2021}
Riello, M., De~Angeli, F., Evans, D.~W., {et~al.} 2021, Astronomy and
  Astrophysics, 649, A3, \dodoi{10.1051/0004-6361/202039587}

\bibitem[{Rizzuto {et~al.}(2020)Rizzuto, Dupuy, Ireland, \&
  Kraus}]{rizzuto2020a}
Rizzuto, A.~C., Dupuy, T.~J., Ireland, M.~J., \& Kraus, A.~L. 2020, The
  Astrophysical Journal, 889, 175, \dodoi{10.3847/1538-4357/ab5aed}

\bibitem[{Rizzuto {et~al.}(2016)Rizzuto, Ireland, Dupuy, \&
  Kraus}]{rizzuto2016}
Rizzuto, A.~C., Ireland, M.~J., Dupuy, T.~J., \& Kraus, A.~L. 2016, The
  Astrophysical Journal, 817, 164, \dodoi{10.3847/0004-637X/817/2/164}

\bibitem[{Roccatagliata {et~al.}(2020)Roccatagliata, Franciosini, Sacco,
  Randich, \& {Sicilia-Aguilar}}]{roccatagliata2020}
Roccatagliata, V., Franciosini, E., Sacco, G.~G., Randich, S., \&
  {Sicilia-Aguilar}, A. 2020, Astronomy and Astrophysics, 638, A85,
  \dodoi{10.1051/0004-6361/201936401}

\bibitem[{Rodrigo \& Solano(2020)}]{rodrigo2020}
Rodrigo, C., \& Solano, E. 2020, The {{SVO Filter Profile Service}}, 182.
\newblock \url{https://ui.adsabs.harvard.edu/abs/2020sea..confE.182R}

\bibitem[{Ryabchikova \& Pakhomov(2015)}]{ryabchikova2015}
Ryabchikova, T., \& Pakhomov, {\relax Yu}. 2015, Baltic Astronomy, 24, 453,
  \dodoi{10.1515/astro-2017-0249}

\bibitem[{Schaefer {et~al.}(2020)Schaefer, Beck, Prato, \&
  Simon}]{schaefer2020a}
Schaefer, G.~H., Beck, T.~L., Prato, L., \& Simon, .~M. 2020, AJ, 160, 35,
  \dodoi{10.3847/1538-3881/ab93be}

\bibitem[{Schaefer {et~al.}(2014)Schaefer, Prato, Simon, \&
  Patience}]{schaefer2014}
Schaefer, G.~H., Prato, L., Simon, M., \& Patience, J. 2014, AJ, 147, 157,
  \dodoi{10.1088/0004-6256/147/6/157}

\bibitem[{Schaefer {et~al.}(2012)Schaefer, Prato, Simon, \&
  Zavala}]{schaefer2012}
Schaefer, G.~H., Prato, L., Simon, M., \& Zavala, R.~T. 2012, The Astrophysical
  Journal, 756, 120, \dodoi{10.1088/0004-637X/756/2/120}

\bibitem[{Schaefer {et~al.}(2006)Schaefer, Simon, Beck, Nelan, \&
  Prato}]{schaefer2006}
Schaefer, G.~H., Simon, M., Beck, T.~L., Nelan, E., \& Prato, L. 2006, The
  Astronomical Journal, 132, 2618, \dodoi{10.1086/508935}

\bibitem[{Service {et~al.}(2016)Service, Lu, Campbell, Sitarski, Ghez, \&
  Anderson}]{service2016}
Service, M., Lu, J.~R., Campbell, R., {et~al.} 2016, Publications of the
  Astronomical Society of the Pacific, 128, 095004,
  \dodoi{10.1088/1538-3873/128/967/095004}

\bibitem[{Simon {et~al.}(1996)Simon, Holfeltz, \& Taff}]{simon1996}
Simon, M., Holfeltz, S.~T., \& Taff, L.~G. 1996, ApJ, 469, 890,
  \dodoi{10.1086/177836}

\bibitem[{Simon {et~al.}(2019)Simon, Guilloteau, Beck, Chapillon, Folco,
  Dutrey, Feiden, Grosso, Pi{\'e}tu, Prato, \& Schaefer}]{simon2019}
Simon, M., Guilloteau, S., Beck, T.~L., {et~al.} 2019, ApJ, 884, 42,
  \dodoi{10.3847/1538-4357/ab3e3b}

\bibitem[{Skrutskie {et~al.}(2006)Skrutskie, Cutri, Stiening, Weinberg,
  Schneider, Carpenter, Beichman, Capps, Chester, Elias, Huchra, Liebert,
  Lonsdale, Monet, Price, Seitzer, Jarrett, Kirkpatrick, Gizis, Howard, Evans,
  Fowler, Fullmer, Hurt, Light, Kopan, Marsh, McCallon, Tam, Van~Dyk, \&
  Wheelock}]{skrutskie2006}
Skrutskie, M.~F., Cutri, R.~M., Stiening, R., {et~al.} 2006, The Astronomical
  Journal, 131, 1163, \dodoi{10.1086/498708}

\bibitem[{Sokal {et~al.}(2018)Sokal, Deen, Mace, Lee, Oh, Kim, Kidder, \&
  Jaffe}]{sokal2018}
Sokal, K.~R., Deen, C.~P., Mace, G.~N., {et~al.} 2018, The Astrophysical
  Journal, 853, 120, \dodoi{10.3847/1538-4357/aaa1e4}

\bibitem[{Sokal {et~al.}(2020)Sokal, {Johns-Krull}, Mace, Nofi, Prato, Lee, \&
  Jaffe}]{sokal2020}
Sokal, K.~R., {Johns-Krull}, C.~M., Mace, G.~N., {et~al.} 2020, ApJ, 888, 116,
  \dodoi{10.3847/1538-4357/ab59d8}

\bibitem[{Stahl {et~al.}(2022)Stahl, {Johns-Krull}, \& Flagg}]{stahl2022}
Stahl, A.~G., {Johns-Krull}, C.~M., \& Flagg, L. 2022, The Astrophysical
  Journal, 941, 101, \dodoi{10.3847/1538-4357/ac8b78}

\bibitem[{Stahl {et~al.}(2021)Stahl, Tang, {Johns-Krull}, Prato, Llama, Mace,
  Joon~Lee, Oh, Luna, \& Jaffe}]{stahl2021}
Stahl, A.~G., Tang, S.-Y., {Johns-Krull}, C.~M., {et~al.} 2021, AJ, 161, 283,
  \dodoi{10.3847/1538-3881/abf5e7}

\bibitem[{{STScI Development Team}(2013)}]{stscidevelopmentteam2013}
{STScI Development Team}. 2013, Astrophysics Source Code Library,
  ascl:1303.023.
\newblock \url{https://ui.adsabs.harvard.edu/abs/2013ascl.soft03023S}

\bibitem[{Sullivan \& Kraus(2022)}]{sullivan2022}
Sullivan, K., \& Kraus, A.~L. 2022, ApJ, 928, 134,
  \dodoi{10.3847/1538-4357/ac5744}

\bibitem[{Tang {et~al.}(2021)Tang, Stahl, {Johns-Krull}, Prato, \&
  Llama}]{tang2021a}
Tang, S.-Y., Stahl, A., {Johns-Krull}, C., Prato, L., \& Llama, J. 2021, JOSS,
  6, 3095, \dodoi{10.21105/joss.03095}

\bibitem[{Tannock {et~al.}(2022)Tannock, Metchev, Hood, Mace, Fortney, Morley,
  Jaffe, \& Lupu}]{tannock2022}
Tannock, M.~E., Metchev, S., Hood, C.~E., {et~al.} 2022, MNRAS, 514, 3160,
  \dodoi{10.1093/mnras/stac1412}

\bibitem[{Torres {et~al.}(2009)Torres, Loinard, Mioduszewski, \&
  Rodr{\'i}guez}]{torres2009a}
Torres, R.~M., Loinard, L., Mioduszewski, A.~J., \& Rodr{\'i}guez, L.~F. 2009,
  The Astrophysical Journal, 698, 242, \dodoi{10.1088/0004-637X/698/1/242}

\bibitem[{Tull {et~al.}(1995)Tull, MacQueen, Sneden, \& Lambert}]{tull1995}
Tull, R.~G., MacQueen, P.~J., Sneden, C., \& Lambert, D.~L. 1995, PASP, 107,
  251, \dodoi{10.1086/133548}

\bibitem[{Valenti(1994)}]{valenti1994}
Valenti, J.~A. 1994, PhD thesis.
\newblock \url{https://ui.adsabs.harvard.edu/abs/1994PhDT........16V}

\bibitem[{Vidotto {et~al.}(2014)Vidotto, Gregory, Jardine, Donati, Petit,
  Morin, Folsom, Bouvier, Cameron, Hussain, Marsden, Waite, Fares, Jeffers, \&
  {do Nascimento}}]{vidotto2014}
Vidotto, A.~A., Gregory, S.~G., Jardine, M., {et~al.} 2014, Monthly Notices of
  the Royal Astronomical Society, 441, 2361, \dodoi{10.1093/mnras/stu728}

\bibitem[{Vrba {et~al.}(1989)Vrba, Rydgren, Chugainov, Shakovskaia, \&
  Weaver}]{vrba1989}
Vrba, F.~J., Rydgren, A.~E., Chugainov, P.~F., Shakovskaia, N.~I., \& Weaver,
  W.~B. 1989, The Astronomical Journal, 97, 483, \dodoi{10.1086/114998}

\bibitem[{Wizinowich {et~al.}(2000)Wizinowich, Acton, Shelton, Stomski,
  Gathright, Ho, Lupton, Tsubota, Lai, Max, Brase, An, Avicola, Olivier, Gavel,
  Macintosh, Ghez, \& Larkin}]{wizinowich2000}
Wizinowich, P., Acton, D.~S., Shelton, C., {et~al.} 2000, Publications of the
  Astronomical Society of the Pacific, 112, 315, \dodoi{10.1086/316543}

\bibitem[{Xiao {et~al.}(2012)Xiao, Covey, Rebull, Charbonneau, Mandushev,
  O'Donovan, Slesnick, \& Lloyd}]{xiao2012}
Xiao, H.~Y., Covey, K.~R., Rebull, L., {et~al.} 2012, The Astrophysical Journal
  Supplement Series, 202, 7, \dodoi{10.1088/0067-0049/202/1/7}

\bibitem[{Yang \& {Johns-Krull}(2011)}]{yang2011}
Yang, H., \& {Johns-Krull}, C.~M. 2011, The Astrophysical Journal, 729, 83,
  \dodoi{10.1088/0004-637X/729/2/83}

\bibitem[{Yang {et~al.}(2005)Yang, {Johns-Krull}, \& Valenti}]{yang2005}
Yang, H., {Johns-Krull}, C.~M., \& Valenti, J.~A. 2005, The Astrophysical
  Journal, 635, 466, \dodoi{10.1086/497070}

\bibitem[{Yuk {et~al.}(2010)Yuk, Jaffe, Barnes, Chun, Park, Lee, Lee, Wang,
  Park, Pak, Strubhar, Deen, Oh, Seo, Pyo, Park, Lacy, Goertz, Rand, \&
  {Gully-Santiago}}]{yuk2010}
Yuk, I.-S., Jaffe, D.~T., Barnes, S., {et~al.} 2010, in {{SPIE Astronomical
  Telescopes}} + {{Instrumentation}}, ed. I.~S. McLean, S.~K. Ramsay, \&
  H.~Takami, {San Diego, California, USA}, 77351M, \dodoi{10.1117/12.856864}

\bibitem[{{Z{\'u}{\~n}iga-Fern{\'a}ndez}
  {et~al.}(2021){Z{\'u}{\~n}iga-Fern{\'a}ndez}, Olofsson, Bayo, Haubois,
  {Corral-Santana}, {Lopera-Mej{\'i}a}, Ronco, Tokovinin, Gallenne, Kennedy, \&
  Berger}]{zuniga-fernandez2021}
{Z{\'u}{\~n}iga-Fern{\'a}ndez}, S., Olofsson, J., Bayo, A., {et~al.} 2021,
  Astronomy and Astrophysics, 655, A15, \dodoi{10.1051/0004-6361/202141985}

\end{thebibliography}
\bibliographystyle{aasjournal}

\end{document}